\documentclass{jpp}
\usepackage{graphicx}
\usepackage{epstopdf, epsfig}
\usepackage{hyperref}
\usepackage{xcolor}

\shorttitle{RTI in HEDP for high Atwood number regime}
\shortauthor{R. K. Bera, Y. Song, and B. Srinivasan}

\title{The effect of viscosity and resistivity on Rayleigh-Taylor instability induced mixing in magnetized high energy density plasmas}

\author{Ratan Kumar Bera\aff{1}
  \corresp{\email{bratan@vt.edu}},
  Yang Song\aff{1},
  \and  Bhuvana Srinivasan\aff{1}
  \corresp{\email{srinbhu@vt.edu}}}

\affiliation{\aff{1}Kevin T. Crofton Department of Aerospace and Ocean Engineering, Virginia Tech, Blacksburg, VA 24060, USA}

\begin{document}

\maketitle

\begin{abstract}
  This work numerically investigates the role of viscosity and
  resistivity on Rayleigh-Taylor instabilities in magnetized
  high-energy-density (HED) plasmas for a high Atwood number and high
  plasma beta regimes surveying across plasma beta and magnetic
  Prandtl numbers.  The numerical simulations are performed using the
  visco-resistive magnetohydrodynamic (MHD) equations.  Results
  presented here show that the inclusion of self-consistent viscosity
  and resistivity in the system drastically changes the growth of the
  Rayleigh-Taylor instability (RTI) as well as modifies its internal
  structure at smaller scales.  It is seen here that the viscosity has
  a stabilizing effect on the RTI. Moreover, the viscosity inhibits
  the development of small scale structures and also modifies the
  morphology of the tip of the RTI spikes. On the other hand, the
  resistivity reduces the magnetic field stabilization supporting the
  development of small scale structures. The morphology of the RTI
  spikes is seen to be unaffected by the presence of resistivity in
  the system.  An additional novelty of this work is in the disparate
  viscosity and resistivity profiles that may exist in HED plasmas and
  their impact on RTI growth, morphology, and the resulting turbulence
  spectra.  Furthermore, this work shows that the dynamics of the
  magnetic field is independent of viscosity and likewise the
  resistivity does not affect the dissipation of enstrophy and kinetic
  energy.  In addition, power law scalings of enstrophy, kinetic
  energy, and magnetic field energy are provided in both injection
  range and inertial sub-range which could be useful for understanding
  RTI induced turbulent mixing in HED laboratory and astrophysical
  plasmas and could aid in interpretation of observations of
  RTI-induced turbulence spectra.
\end{abstract}

\section{Introduction }
\label{sec1}

The Rayleigh-Taylor instability (RTI)
\cite{lord1900investigation,taylor1950instability,
  chandrasekhar1961hydrodynamic}, an important hydrodynamic
instability, occurs at the unstable interface when a high density
fluid is supported by a lower density fluid under the influence of
gravity, or when the interface between two fluids with different
densities is accelerated.  This instability is ubiquitous in nature
and plays an important role in diverse areas of science and
technology, including inertial confinement fusion (ICF)
\cite{Tabak1990,zhou2019,Remington2006,betti1998growth,
  srinivasan2012mechanism,srinivasan2012magnetic,srinivasan2017fluid,srinivasan2019survey,srinivasan2014role,srinivasan2014mitigating,stone2007nonlinear,wang2017theoretical},
astrophysics \cite{Gamezo77,Kifo,Hwang2004,Hester,Loll2013},
geophysics \cite{Boris2007}, and engineering processes
\cite{Lyubimova}.  For instance, the RTI is known to act as an
inhibitor in achieving an ignition grade hot spot in ICF targets
\cite{srinivasan2012mechanism,srinivasan2012magnetic,srinivasan2019survey,srinivasan2014role,srinivasan2014mitigating,ZHOU20171,ZHOU20172}.
RTI occurs in ICF targets during both the acceleration and
deceleration phase of the implosion, leading to undesirable mixing of
hot and cold plasmas.  The RTI is also observed in various
astrophysical phenomena such as supernova explosions and their
remnants (Crab Nebula) \cite{Gamezo77, Kifo, Hwang2004, Hester,
  Loll2013}.  Therefore, a detailed understanding of such
instabilities in high-energy-density (HED) plasmas has implications
for ignition-grade hot-spots, understanding supernova explosions, and
revealing Mega-Gauss (MG) scale magnetic field generation and their
turbulence in astrophysical settings.  The RTI and their mitigation
mechanism in HED plasmas has been thoroughly studied by several
authors experimentally as well as theoretically and numerically
\cite{Remington2006,srinivasan2012mechanism,srinivasan2012magnetic,srinivasan2019survey,
  srinivasan2014role,srinivasan2014mitigating,atzeni2004physics,
  Sun2021,Silveira2017}.  However, there exists a substantial
disagreement between computer simulation results and high-energy
density laboratory experiments or astrophysical observations of the
RTI \cite{Kuranz,modica}. Most of the experiments or astrophysical
observations have noted unusual morphological structure of RTI which
are significantly different from the computer simulation results,
exhibiting strongly suppressed growth of small scale structures and
mass extensions of RT spikes.  This is due to the fact that many
theoretical and numerical studies use conventional hydrodynamic and
magnetohydrodynamic (MHD) depiction where either the self-consistent
effect of magnetic fields, viscosity, and resistivity have been
ignored or they have been considered in isolation.  First observations
of the the magneto-Rayleigh-Taylor instability evolution in the
presence of magnetic and viscous effects have been made in recent
experiments \cite{adams2015observation}.  The impact of magnetic
fields on RTI in the presence of self-consistent viscosity and
resistivity for experimentally and observationally relevant parameter
regimes in HED plasmas remains an open area of research.

The primary purpose of this paper is, therefore, to understand the
role of the viscous and resistive effects on RTI in magnetized HED
plasmas applicable to astrophysical plasmas as well as ICF-based
laboratory experiments. Specifically, this work aims to understand how
RTI dynamics is impacted by varying plasma beta (ratio of thermal
energy to magnetic energy) and magnetic Prandtl number (ratio of
magnetic Reynolds number to Reynolds number). This study focuses on a
high Atwood number and high-$\beta$ regime, where the energy density
in the magnetic field is small compared to the thermal energy in the
fluid.  The Atwood number ($A_t$) is a dimensionless number defined
as, $A_t = (\rho_H -\rho_L)/(\rho_H + \rho_L)$; where $\rho_H$ and
$\rho_L$ represent the mass density of the heavy and light fluid,
respectively.  This distinguishes the current work from previous works
that have examined the role of viscosity and resistivity in isolation
for ICF applications \cite{srinivasan2014mitigating,
  song2020survey}. In addition, this work also presents the evolution
of RTI considering fully varying self-consistent viscosity and
resistivity profiles. To study the RTI dynamics in HED plasmas, the
magnetohydrodynamic (MHD) equations with the inclusion of viscosity
and resistivity are solved in this work.  These visco-resistive MHD
equations are solved in conservation form in 2D (two dimensions) using
the fluid modeling tool PHORCE (Package of High ORder simulations of
Convection diffusion Equations) based on the unstructured
discontinuous Galerkin finite element method
\cite{song2020unstructured, song2021affine, hesthaven2007nodal}. Under
this configuration, simulations have been performed over a wide range
of magnetic Prandtl numbers with the presence of a longitudinal
external magnetic field to reveal the effect of viscosity and
resistivity on the evolution of RTI and magneto-RTI in HED plasmas.
It is observed that the inclusion of viscosity and resistivity
dramatically changes the growth as well as the structures/morphology
of the instability on different length scales.  It is seen here that
the presence of viscosity stabilizes the growth of the RTI and
modifies the morphology of the tip of RTI fingers, inhibiting the
traditional mushroom cap structures. On the other hand, the morphology
of the RTI spikes is found to be independent of resistivity. The
presence of resistivity assists in the development of small scale
structures by reducing the magnetic field stabilization.  When
considering spatially-varying viscosity and resistivity with highly
disparate profiles, there is a significant impact on the RTI evolution
in the high Atwood number regime studied in this work.  In this paper,
the numerical growth rates of RTI obtained from the simulations are
compared with their corresponding analytical values obtained from
linear theory.  Furthermore, it is also seen here that the dynamics of
magnetic field is independent of viscosity and likewise the
resistivity does not affect the dynamics of enstrophy and kinetic
energy.  In addition, this work presents the power law scaling of
enstrophy, kinetic energy, and magnetic field energy in both the
injection range and inertial sub-range of power spectra for different
viscosity and resistivity cases, which could be useful for
understanding the RTI induced turbulent mixing in HED plasmas.

The manuscript has been organized as follows. In Section \ref{sec2}, a
brief description of the governing equations is presented to study the
RTI process in magnetized HED plasmas.  Section \ref{sec3} discusses
the simulation techniques and problem setup for the study.  Section
\ref{sec4} presents the simulation results, comparison with theory,
and discussions.  Section \ref{sec5} presents the summary and
conclusion.

\section{Governing Equations}
\label{sec2}
In this section, the basic governing equations are presented for the
study of RTI in magnetized HED plasmas in the presence of an applied
horizontal magnetic field, viscosity, and resistivity. Thermal
conduction is neglected in this study to focus on the impact of
viscosity, resistivity, and magnetic fields.  The generalized Lagrange
multiplier-magnetohydrodynamic (GLM-MHD) equations
\cite{munz2001godunov,dedner2002hyperbolic} with the inclusion of
viscosity and resistivity are solved.  The compressible MHD equations
are given by,
\begin{equation}    
  \frac{\partial \rho}{\partial t} + \mathbf{\nabla} \cdot (\rho \mathbf{u} )  = 0 
  \label{eq1}
\end{equation}

\begin{equation}
  \frac{\partial \rho  \mathbf{u} }{\partial t} + \nabla \cdot \left( \rho \mathbf{u} \mathbf{u} + p\mathbf{I} {-\frac{\mathbf{B}\mathbf{B}}{\mu_0}+ \frac{B^2}{2\mu_0}\mathbf{I} }\right)  = {- \rho \mathbf{g}}  + \nabla \cdot \mathbf{\pi} 
  \label{eq2}
\end{equation}

\begin{equation}
  \frac{\partial \epsilon}{\partial t} + \nabla \cdot \left[ \left(\epsilon + p { + \frac{B^2}{2\mu_0}} \right)\mathbf{u} {-\frac{(\mathbf{B} \cdot \mathbf{u})}{\mu_0} \mathbf{B} } \right]  
  = {- \rho \mathbf{g} \cdot \mathbf{u}} + \nabla \cdot (\mathbf{u} \mathbf{\pi})  -\frac{1}{\mu_0} \nabla \cdot (\frac{\eta}{\mu_0} \nabla \times \mathbf{B})
  \label{eq3}
\end{equation}

\begin{equation}
  {\frac{\partial \mathbf{B}}{\partial t}+\nabla \cdot (\mathbf{u}\mathbf{B}-\mathbf{B}\mathbf{u})} + \nabla \psi = -\frac{1}{\mu_0}\nabla \times (\eta \nabla \times \mathbf{B})\\
  \label{eq4}
\end{equation}

\begin{equation}
  \frac{\partial \psi}{\partial t} + C_h^2 \nabla \cdot \mathbf{B} = - \frac{C_{h}^2}{C_{p}^2}\psi
  \label{eq5}
\end{equation}
where $\rho$, $\mathbf{u}$, $p$, $\mathbf{g}$ and $\mathbf{B}$
represent the mass density, fluid velocity, pressure, gravitational
field, and magnetic field, respectively. Here $\epsilon = p/(\gamma
-1) +\rho u^2 /2 + B^2/2\mu_0$ defines the total energy; where
$\gamma$ is the ratio of specific heats, and is normally taken as
$5/3$ for monatomic gases assuming an ideal gas law.  Here $p$ is the
pressure.  For the equation of state, an ideal gas law $p= (Z_i+1)\rho
k_B T_i/m_i$ is assumed; where $Z_i$, $m_i$, $k_b$, and $T_i$
represent the charge state of ion, mass of the ion, Boltzmann
constant, and temperature of the ion, respectively.  Here $\psi$,
$C_h$, and $C_p$ represent the divergence cleaning potential,
hyperbolic cleaning speed, and parabolic cleaning speed, respectively.
A user-specified parameter $C_r = C_p^2/C_h^2 $ is defined to
determine the ratio between hyperbolic and parabolic divergence
cleaning. If $C_r$ is very large, the divergence error will only be
transported through the hyperbolic term. $C_h$ is calculated based on
the grid sizes and CFL number \citep{dedner2002hyperbolic}.  In the
simulations presented here, $C_r =99999$ is set to be very large so
that only hyperbolic cleaning dominates.  In the above equations
$\mathbf{\pi}$ and $\eta$ represent the viscous stress tensor and
electrical resistivity coefficient, respectively.  In this study, the
Braginskii formulation \cite{braginskii1965transport} for calculating
viscosity and resistivity co-efficient is used, $\mu = 0.96n_i k_B T_i
\tau_i $ and $ \eta =m_e/1.96n_e q_e^2 \tau_e$, where $\tau_e$ and
$\tau_i$ are the collision times for electron and ion, respectively.
Note that the viscosity and resistivity can also be presented in terms
of Reynolds ($Re$) and magnetic Reynolds number ($Re_m$) defined as,
$Re=\rho VL/\mu $ and $Re_m=\mu_0 VL/\eta $; where $V$ and $L$
represent some reference velocity and length, respectively.

\section{Numerical simulation and problem setup}
\label{sec3}
This section presents the simulation techniques and problem setup used
for studying the role of viscosity and resistivity on RTI in
magnetized HED plasmas. The simulations presented here are in planar
geometry and in 2D.  A significant amount of insight can be gained
from 2D studies particularly where observations may be dominated by 2D
evolution of perturbation growth.  In other words, this is true when
the wavelength of perturbation for RTI growth in the considered
directions is much smaller than the wavelength of perturbation in the
third direction. This approximation would be particularly well-suited
for cases where magnetic fields influence RTI growth leading to
regimes where the perturbation growth are more ``2D-like''.  Most of
the past literature on 2D MHD turbulence, not specific to RTI, has
focused on incompressible MHD models \cite{orszag1979small,
  biskamp2001two} whereas this work uses a compressible MHD model with
a focus on the evolution of the RTI.  However, a fully 3-D RTI
turbulence study would be important to understand the RTI induced
turbulence accounting for 3-D perturbations and this would constitute
future studies.  In this paper, the code PHORCE (Package of High ORder
simulations of Convection diffusion Equations)
\cite{song2020unstructured,song2021affine} developed at Virginia Tech
is used for the 2D RTI study. PHORCE is based on the nodal
unstructured discontinuous Galerkin method \cite{hesthaven2007nodal}
and solves fluid equations (\ref{eq1}-\ref{eq5}) in conservation
form. To advance the simulation in time, an explicit fourth-order
five-stage strong stability-preserving Runge-Kutta (SSP-RK)
\cite{song2020unstructured} scheme has been implemented.  Several
limiters and filters are applied in PHORCE to preserve the positivity
of density and pressure and to diffuse the numerical oscillations that
typically occur due to strong discontinuities. The code uses an affine
reconstructed discontinuous Galerkin (aRDG) scheme
\cite{song2020unstructured,song2021affine} to solve the diffusion
terms and to self-consistently capture the effect of spatially varying
Reynolds numbers (viscous effects) and magnetic Reynolds numbers
(resistive effects).

The RTI simulations have been performed in a rectangular domain with
$x\in [-L_x/2, L_x/2]$, $y \in [-L_y/2, L_y/2]$; where $L_x$ and $L_y$
represent the width and height of the simulation domain, respectively.
The simulations are performed with $2000 \times 1000$ cells.  The
gravitational field $\mathbf{g}=-g\hat{y}$.  The simulations are
performed using ``conducting wall'' boundary conditions along the $y$-
direction and ``periodic'' boundary condition along the $x$-direction.
In equilibrium, the simulation is initialized using the standard
hyperbolic tangent density profile given by,
\begin{equation}
  \rho=\frac{(\rho_H -\rho_L)}{2}[\tanh{(\alpha y/L_y)} +1]+ \rho_L.
\end{equation}
 In the above equation, $\alpha$ defines the width of the hyperbolic
 tangent function.  In the simulations presented here $\alpha$ is
 taken to be $0.01$ in order to provide a sharp gradient at the
 interface.  The pressure profile is initialized as,
\begin{equation}
  p=p_0 - \frac{(\rho_H +\rho_L)}{2}g y-g\frac{(\rho_H
    -\rho_L)}{2}\frac{L_y}{\alpha} \ln{\cosh{(\alpha y/L_y)}}
\end{equation}   
where $p_0$ represents the background pressure of the system.  To
excite the multimode RTI in the simulation, the $y$-component of
velocity at the interface ($y=0$ plane) is perturbed as, $v=
\Sigma_{m=1}^{40} 0.01 R(m) cos (2 \pi (m x/Lx + R(m)))exp(-\xi y^2)$
at $t=0$; where $R(m)$ and $\xi$ represent the random number generator
function of $m$ random numbers and the spatial width along the $y$
direction over which the perturbation falls at the interface, with
$\xi=1000$.

In this work, all the simulation results are presented in normalized
units. The following normalization factors have been used, $x
\rightarrow \frac{x}{L_x}$, $y \rightarrow \frac{y}{L_x}$, $t
\rightarrow t\gamma_{RT}$, $\rho \rightarrow \frac{\rho}{\rho_L}$, $g
\rightarrow \frac{g}{\gamma_{RT}^2 L_x}$ and $k_x \rightarrow k_x
L_x$. Here $\gamma_{RT} =\sqrt{A_t g k}$, represents the growth rate
of the RTI associated with the wave number $k=2 \pi/\lambda$; where
$\lambda$ is the wavelength of the mode
\cite{chandrasekhar1961hydrodynamic}.  As the simulations have been
conducted with multimode perturbations having mode number $m = 1-40$,
note that the value of $\gamma_{RT}$ would be different for different
modes (or wavelengths). The growth rate becomes maximum for smallest
wavelength and minimum for longest wavelength modes. To calculate the
the value of $\gamma_{RT}$ for the normalization of time, the smallest
mode of perturbation ($m=40$) having wavelength $\lambda= L_x/40$ has
been selected.

In some flows in HED plasmas, such as in ICF and supernovae explosions
\cite{sauppe2019, Burton, Dimonte2005, Cabot,
  srinivasan2014mitigating, srinivasan2012magnetic}, the Atwood number
can reach a very high value ($A_t\geq 0.85$) and the temperature can
have a large variation in the domain.  As a result, a large variation
in Reynolds and magnetic Reynolds numbers may exist in the domain.  In
this work, the plasma parameters are selected to access highly varying
density and temperature regimes in laboratory and astrophysical
plasmas where the viscosity and resistivity may be important.  The
parameters are summarized in Table~\ref{Tab1} in normalized form.  The
simulations use an initial plasma beta $\beta^{ini} =2 \mu_0
p_0/{B_x^{ext }}^2 = 5000$ whenever an external horizontal magnetic
field ($B_x^{ext }$) exists in the system.

\begin{table}
  \begin{center}
    \def~{\hphantom{0}}
    \begin{tabular}{lccc}
      Parameter & &  values \\ [3pt]
      \hline
      Atwood number ($A_t$) &  & 0.95866 \\
      Light fluid density ($\rho_L$) &  & 1 \\
      Heavy fluid density ($\rho_H$) &   & 47  \\
      gravitational acceleration ($g$) &  & $ 4.2 \times 10^{-3}$ \\
      Initial plasma beta ($\beta^{ini}$) & & $\approx 5000$ \\
    \end{tabular}
    \caption{Summary of plasma parameters in normalized form.}
    \label{Tab1}
  \end{center}
\end{table}

\begin{figure}
  \centering
  \includegraphics[width=0.8\linewidth]{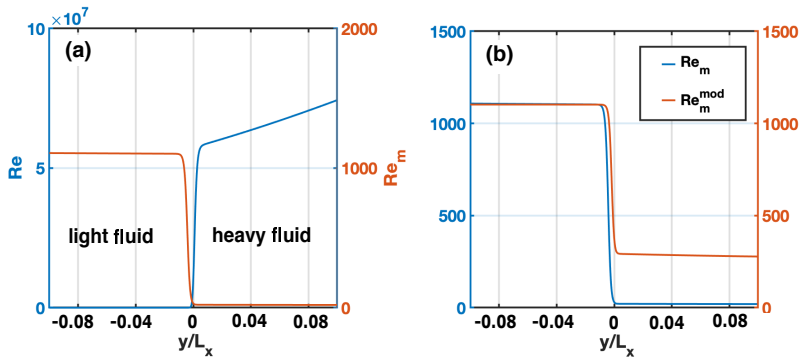}
  \caption{The left plot (a) shows the profile of Reynolds number ($Re$) and magnetic Reynolds number ($Re_m$) as function of vertical height ($y/L_x$) in the domain. The right plot (b) shows the profile of magnetic Reynolds number ($Re_m$) along with modified magnetic Reynolds number ($Re_m^{mod}$) profile as function of vertical height ($y/L_x$).}
  \label{fig1}
\end{figure}

Using the parameters given in the Table~\ref{Tab1} and using the
expressions for isotropic viscosity ($\mu$) and resistivity ($\eta$)
mentioned in Section~\ref{sec2}, the Reynolds number $Re=\rho V_{RT}
L_y/\mu$ and magnetic Reynolds number $Re_m=\mu_0 V_{RT} L_y/\eta$,
are plotted as a function of vertical height ($y/L_x$) in Fig
\ref{fig1}(a).  Here $V_{RT}=L_y \gamma_{RT} $ defines the terminal
velocity of the RTI.  
Note that $Re$ and $Re_m$ are in the range of $485-7.3 \times 10^7$
and $ 20 -1105$, respectively.  The profile of resistivity (and
corresponding magnetic Reynolds number) has been modified to ensure
the resistive time step is larger than the hyperbolic time step since
an explicit time-stepping scheme is used in this work.  The following
form of modified resistivity ($\eta_{mod}$) has been used, $\eta_{mod}
= \eta /a + b$ ; where $a = 18.5$ and $b =7.3 \times 10^{-9}$ are
constants.  Using the modified expression of resistivity $\eta^{mod}$,
the modified profile of magnetic Reynolds number ($Re_m^{mod}$) is
plotted in Fig.~\ref{fig1}(b). Note that the resistivity profile is
modified in the heavy fluid to increase the minimum value of the
magnetic Reynolds number from $20$ to $285$.  For the simulations
presented here, the modified resistivity profile has been used to
capture the essential physics of RTI in the presence of resistivity.
The magnetic Prandtl number, $Pr_m = Re_m /Re =\nu/\eta$ (where $\nu
=\mu/\rho$ is the kinematic viscosity), is a dimensionless quantity
that estimates the ratio of momentum and magnetic diffusivity. In
Fig. 1b, $Pr_m$ varies from $2$ for $y/L_x<0$ to $4\times10^{-6}$ for
$y/L_x>0$ producing a significant variation across the domain.

\section{Simulation results and discussion}
\label{sec4}
The simulations have been performed for different values of magnetic
Prandtl numbers to elucidate the role of viscosity and resistivity on
the Rayleigh-Taylor and magneto-Rayleigh-Taylor instability.
Table~\ref{Tab2} summarizes all simulation cases performed here for
different values of plasma beta (external magnetic field) and magnetic
Prandtl numbers (Reynolds numbers and magnetic Reynolds numbers).
This section discusses the results and findings of each case that is
presented.

\begin{table}
  \begin{center}
    \begin{tabular}{cccccccc}
      runs & &$\beta^{ini}$ & $Re$ &$Re_m$& $Pr_m$\\
      \hline
      run-1& & $\infty$ & $\infty$ & $0$\\
      run-2 && 5000 & $\infty$ & $\infty$\\
      &  &  &\\
      run-3 & & $\infty$ & $2 \times 10^3$ & $\infty$ & $\infty$\\
      run-4& &$\infty$ & $2 \times 10^6$ & $\infty$ & $\infty$\\
      run-5& &5000& $2 \times 10^3$ & $\infty$ & $\infty$\\
      run-6& &5000 & $2 \times 10^6$ & $\infty$ & $\infty$\\
      &  &  &\\
      run-7& & 5000 & $\infty$ & 285 & 0\\
      run-8& &5000 & $\infty$ & 1105 & 0\\
      &  &  &\\
      run-9&  &5000 & $2 \times 10^3$& 285 & $0.1$\\
      run-10&  &5000 & $2 \times 10^3$ & 1105 & $0.5$\\
      run-11 & &5000 & $2 \times 10^6$ & 285 & $1\times 10^{-4}$\\
      run-12 & & 5000 & $2 \times 10^6$ & 1105 & $5\times 10^{-4}$\\
      &  &  &\\
      run-13& &$\infty$ & fully varying & $\infty$ & $\infty$ \\
      run-14& &5000 & fully varying & $\infty$ & $\infty$\\
      &  &  &\\
      run-15& &5000 & fully varying & 285 & $0.5-4\times10^{-6}$\\
      run-16& &5000 & fully varying & 1105 & $2-1.5\times10^{-5}$\\
      &  &  &\\
      run-17& &5000 & fully varying & fully varying &$2-4\times10^{-6}$\\
    \end{tabular}
    \caption{Summary of numerical simulations performed here.}
    \label{Tab2}
  \end{center}
\end{table}


\subsection{Simulation results for inviscid, irresitive cases: run-1 and run-2} \label{sec:inviscirres}

\begin{figure*}
  \centering
  \includegraphics[width=\linewidth]{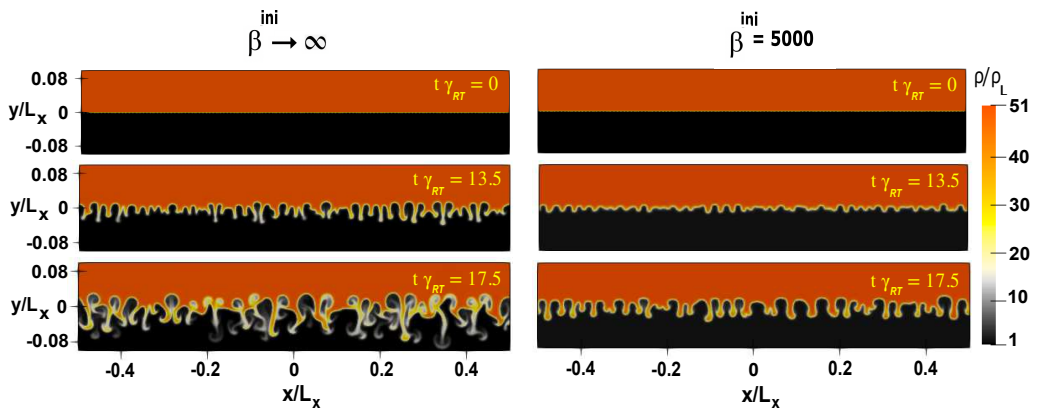}
  \caption{Plot of mass density ($\rho/\rho_L$) profiles at different
    times for $\beta^{ini} \rightarrow \infty$ (left) and
    $\beta^{ini}=5000$ (right); where $\mu=0$ and $\eta=0$.  Note the
    stabilizing effect of an applied horizontal magnetic field on
    overall RTI and the damping of short-wavelength modes. }
  \label{fig2}
\end{figure*}

\begin{figure}
  \centering
  \includegraphics[width=0.6\linewidth]{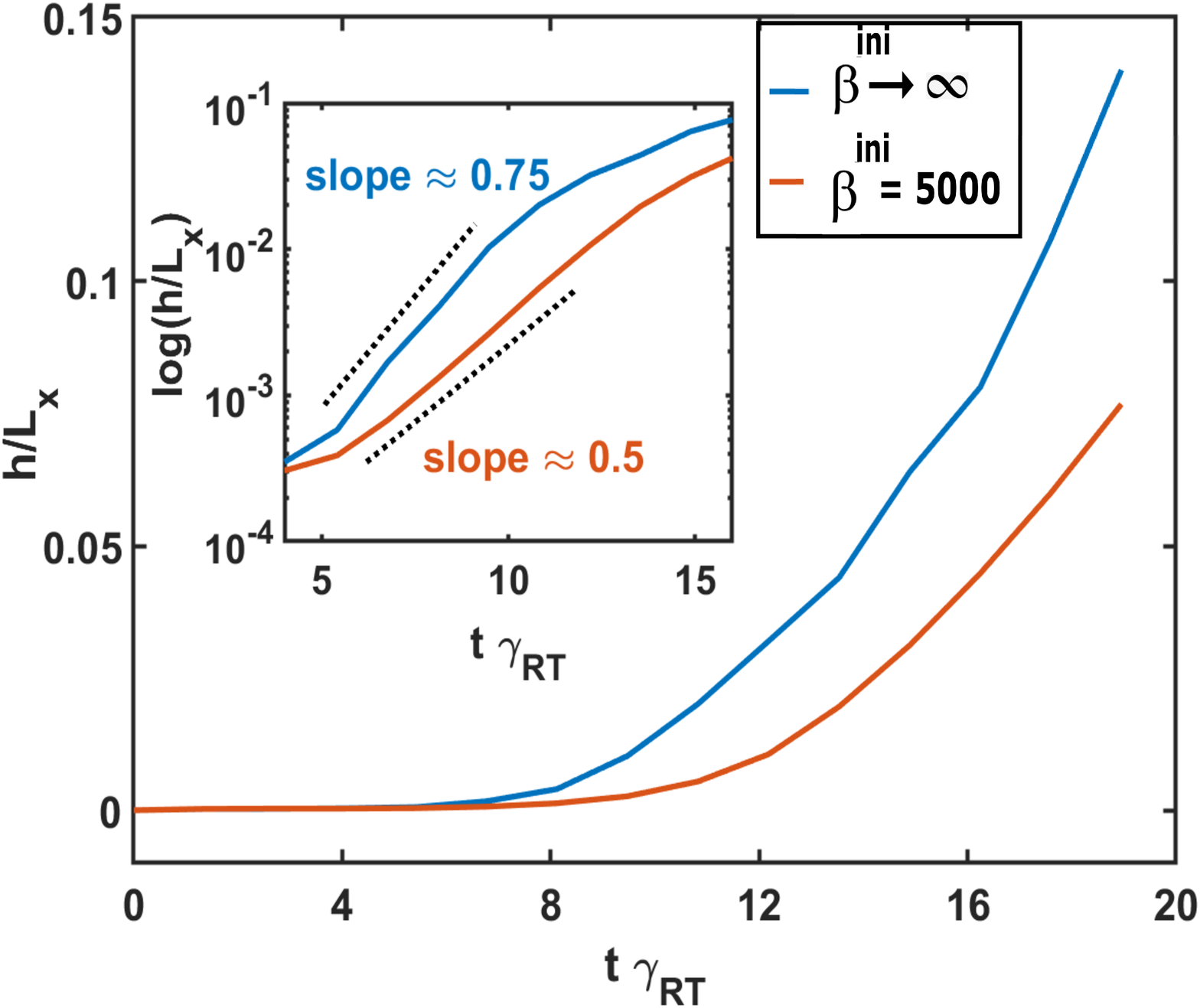}
  \caption{Plot of peak bubble-to-spike distance ($h/L_x$) over time
    ($t \gamma_{RT}$) for $\beta^{ini} \rightarrow \infty$ and
    $\beta^{ini}=5000$; where $\mu=0$ and $\eta=0$ to estimate a
    numerical growth rate.  Note the growth rate is reduced with an
    applied horizontal magnetic field as expected.}
  \label{fig3}
\end{figure}

Simulations for inviscid ($\mu=0$) and irresitive ($\eta=0$) cases are
performed (see run-1 and run-2 in Table~\ref{Tab2}).  Fig.~\ref{fig2}
presents plots of mass density ($\rho/\rho_L$) at different times for
$\beta^{ini} \rightarrow \infty$ (no initial external horizontal
magnetic field) and for $\beta^{ini} =5000$ (in the presence of
initial external horizontal magnetic field).  As expected, the height
of the RTI mixing region or the height of the RTI fingers reduces in
the presence of an applied horizontal magnetic field.  Note ths
suppression of small-scale structures due to the presence of the
magnetic field.  To calculate the growth rate, the peak
bubble-to-spike distance ($h/L_x$) over the normalized times ($t
\gamma_{RT}$) for both $\beta^{ini} \rightarrow \infty$ and
$\beta^{ini} =5000$ is presented in Fig.~\ref{fig3}. In the
simulations, the height has been measured by tracking the difference
of the upper and lower boundary of the RTI mixing region.  As shown in
the subplot of Fig.~\ref{fig3}, the numerical growth rates are
calculated from the slope of the plot $log (h/L_x)$ vs. $t
\gamma_{RT}$.  The numerical growth rate obtained from the simulations
for both $\beta^{ini} \rightarrow \infty$ and $\beta^{ini}=5000$ are
$0.75 \gamma_{RT}$ and $0.5 \gamma_{RT}$, respectively.  The growth of
RTI significantly decreases in the presence of applied horizontal
magnetic field as expected.  The analytical expression of growth rate
($\gamma_{RT}$) of RTI for purely hydrodynamic flows (no viscosity, no
resistivity, and no magnetic field) is given by
\cite{chandrasekhar1961hydrodynamic} as,
\begin{equation}
  \gamma_{RT} =\sqrt{A_t g k} .
\end{equation}
Using the parameters given in Table~\ref{Tab1} and $k= 80\pi/L_x$ (for
$\lambda =L_x/40$), the analytical values of the growth rate
$\gamma_{RT}$ can be estimated as $2.69 \times 10^9$s$^{-1}$ for a
single mode that is estimated to be the fastest growing early in time.
The numerical growth rate is $0.75 \gamma_{RT}= 2 \times 10^9$s$^{-1}$
but this is for a multimode growth rate which explains the
difference between the analytical and numerical values.  As time
evolves, the nonlinear interactions between modes significantly
changes the dominant wave number.  When an applied magnetic field
$\textbf{B}^{ext}$ exists, the RTI growth rate becomes
\cite{chandrasekhar1961hydrodynamic,Jun1996},
\begin{equation}
  \gamma_{RT}^B =\sqrt{A_t gk - \frac{(\textbf{B} \cdot \textbf{k})^2}{2 \pi \mu_0 (\rho_H +\rho_L)}}.
\end{equation}
Note that the RTI is affected by the horizontal magnetic field ($
\textbf{B} \parallel \textbf{k}$) and is not directly impacted by
magnetic fields that are normal to the interface when using an MHD
model. In Fig.~\ref{fig2} for $\beta^{ini}=5000$, the height of the
mixing region is decreased along with suppression of the small
structures. In this case, one can approximately calculate the
wavelength of RTI fingers by calculating the number of RTI fingers in
the domain. This technique suggests approximately $30$ RTI spikes
growing at this time. Therefore, the effective smallest wavelength is
approximately $\approx L_x/30$.  When an appropriately aligned
magnetic field is initialized, the value of the peak magnetic field in
the system increases with time as RTI grows.  For example, the plasma
$\beta$ becomes $226$ from an initial value of $5000$ at time
$t\gamma_{RT} =13.5$.  Using the parameters given in Table~\ref{Tab1},
$\beta=226$, and $k_x L_x= 60\pi$, the analytical values of the growth
rate $\gamma_{RT}^B$ can be estimated as $\gamma_{RT}^B =0.63
\gamma_{RT}$. The numerical growth rate obtained from the simulation
shows good agreement with the analytical value for $\beta^{ini}=5000$
considering that these are estimates for multimode simulations.

\begin{figure}
  \centering
  \includegraphics[width=0.8\linewidth]{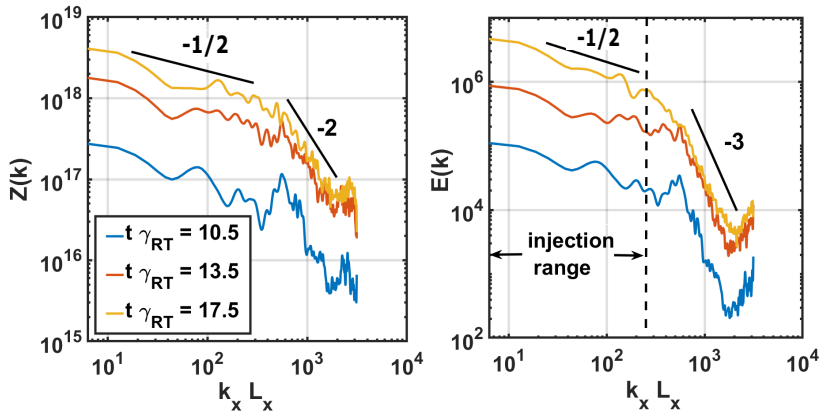}
  \caption{Evolution of enstrophy ($Z$) and kinetic energy ($E$)
    spectra as a function of wave number ($k_x L_x$) for $\beta^{ini}
    \rightarrow \infty$; where $\mu=0$ and $\eta=0$. }
  \label{fig4}
\end{figure}

The enstrophy ($Z$), kinetic energy ($E$), and magnetic field energy
($B^2$) averaged over the vertical direction ($y$) of the system is
defined as,
\begin{equation}
  Z= \langle \omega^2 \rangle =\int_{-L_y/2}^{L_y/2} \omega^2 dy ; \hspace{0.5cm} E=\frac{1}{2}\langle u^2 \rangle; \hspace{0.5cm} B^2=\langle B^2 \rangle
\end{equation} 
where $\mathbf{\omega} = \nabla \times \mathbf{u}$ represents the
fluid vorticity. In 2D mixing and turbulence, the enstrophy ($Z$),
kinetic energy ($E$), and magnetic field energy ($B^2$) are important
quantities as they appear to be the only quadratic constants of
motion.  In Fig.~\ref{fig4}, the evolution of enstrophy ($Z$) and
kinetic energy ($E$) spectra are presented as a function of normalized
wave number $k_x L_x$ at different times for $\beta^{ini} \rightarrow
\infty$. Note that there will be no magnetic field for $\beta^{ini}
\rightarrow \infty$.

The spectra can be separated into three regions based on the range of
$k_x L_x$. The first region with $k_x L_x \leq 80 \pi$ is known as the
injection range where all the external perturbation modes exist. All
the energy has been injected to the system within these
wavelengths. The second region $80\pi \leq k_x L_x \leq 600\pi$ or the
middle range is the inertial sub-range. This is the regime which
basically connects the injection range to the dissipation range. The
third region where $ k_x L_x \geq 600\pi$ is the dissipation range
which accounts for grid scales as $L_x=2000 \Delta x$; where $\Delta
x$ is the grid size along the $x$-direction. As physical dissipation
(viscosity and resistivity) is absent in the system for the
simulations in this section, the only dissipation mechanism is,
therefore, governed by the numerical dissipation. All the energy for
modes smaller than or equal to the grid size is dissipated by
numerical dissipation.  For $\beta^{ini} \rightarrow \infty$ (see
Fig.~\ref{fig4}), note that the enstrophy ($Z$) and kinetic energy
($E$) increase equally in all the available modes in the system with
time as long as $t\gamma_{RT} \leq 13.5$.  At $t \gamma_{RT}= 17.5$,
the transfer of kinetic energy as well as enstrophy is seen from short
wavelength modes to long wavelength modes. This happens due to the
nonlinear interactions of the modes leading to the formation of longer
wavelength modes with time. As a result, the small scale structures
get modified changing the growth rate in the nonlinear regime for
$\beta^{ini} \rightarrow \infty$.  The numerically obtained power law
scalings for the enstrophy, kinetic energy, and magnetic field energy
spectra at both the injection and inertial sub-range are included in
Fig.~\ref{fig4}.  In this case, the spectra of kinetic energy and
enstrophy obey the following power scaling laws in the injection range
($k_x L_x \leq 80 \pi$),
\begin{equation}
  E(k) \sim k_x^{-1/2}
\end{equation}
\begin{equation}
  Z(k) \sim k_x^{-1/2}.
\end{equation}
In the inertial sub-range ($80\pi \leq k_x L_x \leq 600\pi$), the
spectra are found to obey different power laws,
\begin{equation}
  E(k) \sim k_x^{-3}
\end{equation}
\begin{equation}
  Z(k) \sim k_x^{-2}.
\end{equation}

\begin{figure*}
  \centering
  \includegraphics[width=\linewidth]{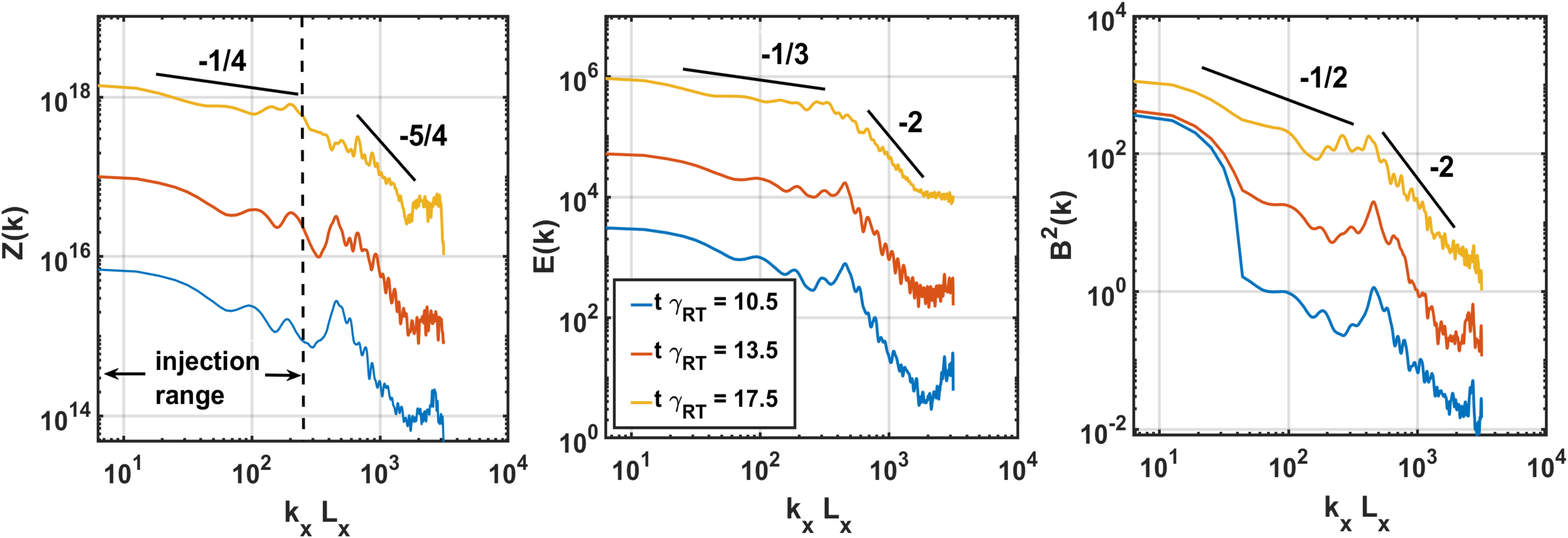}
  \caption{Evolution of enstrophy ($Z$), kinetic energy ($E$), magnetic field energy ($B^2$) spectra as a function of wave number ($k_x L_x$) for  $\beta^{ini}=5000$; where $\mu=0$ and $\eta=0$.}
  \label{fig5}
\end{figure*}
For $\beta^{ini}=5000$, the evolution of enstrophy ($Z$), kinetic
energy ($E$), and magnetic field energy ($B^2$) spectra as a function
of wave number $k_x L_x$ at different times is shown in
Fig.~\ref{fig5}.  The enstrophy ($Z$), kinetic energy ($E$), and
magnetic field energy ($B^2$) increase equally in all modes in the
system until $t \gamma_{RT} = 17.5$ for $\beta^{ini}=5000$. There is
no transfer of kinetic energy, enstrophy, and magnetic field energy
over the modes. This is because the spectrum still lies in the linear
regime due to the presence of a horizontal magnetic field. The
magnetic field opposes the growth of the RTI and decreases the
vertical velocity of the fluid. In this case, the spectra of kinetic
energy, enstrophy, and magnetic field energy, obtained from the
numerical simulations, obey the following power laws in the injection
range ($k_x L_x \leq 80 \pi$)
\begin{equation}
  E(k) \sim k_x^{-1/3}
\end{equation}
\begin{equation}
  Z(k) \sim k_x^{-1/4}
\end{equation}
\begin{equation}
  B^2(k) \sim k_x^{-1/2}.
\end{equation}
Similarly, the power law in the inertial sub-range ($80\pi \leq k_x
L_x \leq 600\pi$) for $\beta^{ini}=5000$ is found to be
\begin{equation}
  E(k) \sim k_x^{-2}
\end{equation}
\begin{equation}
  Z(k) \sim k_x^{-5/4}
\end{equation}
\begin{equation}
  B^2(k) \sim k_x^{-2}.
\end{equation}
The slope of the spectra in the inertial sub-range decreases with the
presence of a horizontal magnetic field. The slope of the inertial
sub-range measures the rate at which the energy is transferred from
large scale to small scales or vice versa. In other words, it defines
the rate at which the larger scales get fragmented into smaller scales
and vice versa due to mixing. Therefore, this shows that the rate of
small scale formation due to RTI mixing decreases with the application
of horizontal magnetic field.  The scaling of these power laws in both
injection and inertial sub-range for these cases (run-1-2) have been
summarized in Tables~\ref{Tab3} and \ref{Tab4}.  Note that the
numerical dissipation is active in the range of $ k_x L_x > 600\pi$.
As a result, all the energy is also seen to grow proportionally with
time in this regime.

\begin{table}
  \begin{center}
    \begin{tabular}{cccccccc}
      runs & injection range power law& \\
      \hline
      run-1 & $Z(k) \sim k_x^{-1/2}$, $E(k) \sim k_x^{-1/2}$  &\\
      run-2 & $Z(k) \sim k_x^{-1/4}$, $E(k) \sim k_x^{-1/3}$, $B^2(k) \sim k_x^{-1/2}$ & \\
      &    &\\
      run-3 & $Z(k) \sim k_x^{-1/4}$, $E(k) \sim k_x^{-1/4}$  &\\
      run-4& $Z(k) \sim k_x^{-1/5}$, $E(k) \sim k_x^{-1/2}$, $B^2(k) \sim k_x^{-1}$ & \\
      run-5& $Z(k) \sim k_x^{-1/2}$, $E(k) \sim k_x^{-1/2}$  & \\
      run-6&  $Z(k) \sim k_x^{-1/2}$, $E(k) \sim k_x^{-1/2}$, $B^2(k) \sim k_x^{-1/2}$ & \\
      &    &\\
      run-7 & $Z(k) \sim k_x^{-1/2}$, $E(k) \sim k_x^{-1/2}$, $B^2(k) \sim k_x^{-1}$ &  \\
      run-8 & $Z(k) \sim k_x^{-0.3}$, $E(k) \sim k_x^{-1/2}$, $B^2(k) \sim k_x^{-1}$ & \\
      &   &\\
      run-9&  $Z(k) \sim k_x^{-1/2}$, $E(k) \sim k_x^{-1/2}$, $B^2(k) \sim k_x^{-1/2}$ &\\
      run-10&   $Z(k) \sim k_x^{-1/4}$, $E(k) \sim k_x^{-0.4}$, $B^2(k) \sim k_x^{-1/2}$  & \\
      run-11 &  $Z(k) \sim k_x^{-1/2}$, $E(k) \sim k_x^{-1/2}$, $B^2(k) \sim k_x^{-1}$  & \\
      run-12 &  $Z(k) \sim k_x^{-1/2}$, $E(k) \sim k_x^{-1/2}$, $B^2(k) \sim k_x^{-1}$  & \\
      &    &\\
      run-13&  $Z(k) \sim k_x^{-0.3}$, $E(k) \sim k_x^{-1/2}$  &  \\
      run-14 & $Z(k) \sim k_x^{-0.3}$, $E(k) \sim k_x^{-0.3}$, $B^2(k) \sim k_x^{-1/2}$  & \\
      &  & &\\
      run-15 & $Z(k) \sim k_x^{-0.3}$, $E(k) \sim k_x^{-1/2}$, $B^2(k) \sim k_x^{-1/2}$  & \\
      run-16& $Z(k) \sim k_x^{-1/2}$, $E(k) \sim k_x^{-1/2}$, $B^2(k) \sim k_x^{-1/2}$  & \\
      &  & &\\
      run-17 & $Z(k) \sim k_x^{-1/2}$, $E(k) \sim k_x^{-1/2}$, $B^2(k) \sim k_x^{-0.3}$  & \\
    \end{tabular}
    \caption{Summary of power laws for the numerical simulations in
      injection range.}
    \label{Tab3}
  \end{center}
\end{table}

\begin{table}
  \begin{center}
    \begin{tabular}{cccccccc}
      runs & inertial sub-range power law&\\
      \hline
      run-1 & $Z(k) \sim k_x^{-2}$, $E(k) \sim k_x^{-3}$  &\\
      run-2 & $Z(k) \sim k_x^{-5/4}$, $E(k) \sim k_x^{-2}$, $B^2(k) \sim k_x^{-2}$ & \\
      &    & \\
      run-3 & $Z(k) \sim k_x^{-5/2}$, $E(k) \sim k_x^{-4}$  &\\
      run-4& $Z(k) \sim k_x^{-3}$, $E(k) \sim k_x^{-4}$, $B^2(k) \sim k_x^{-3}$ & \\
      run-5& $Z(k) \sim k_x^{-5/2}$, $E(k) \sim k_x^{-3}$  & \\
      run-6&  $Z(k) \sim k_x^{-5/2}$, $E(k) \sim k_x^{-5/2}$, $B^2(k) \sim k_x^{-3}$ & \\
      &    &\\
      run-7 & $Z(k) \sim k_x^{-2}$, $E(k) \sim k_x^{-3}$, $B^2(k) \sim k_x^{-4}$ &  \\
      run-8 & $Z(k) \sim k_x^{-5/2}$, $E(k) \sim k_x^{-4}$, $B^2(k) \sim k_x^{-4}$ & \\
      &   &\\
      run-9&  $Z(k) \sim k_x^{-5/2}$, $E(k) \sim k_x^{-7/2}$, $B^2(k) \sim k_x^{-5}$ &\\
      run-10&   $Z(k) \sim k_x^{-2}$, $E(k) \sim k_x^{-4}$, $B^2(k) \sim k_x^{-4}$  & \\
      run-11 &  $Z(k) \sim k_x^{-2}$, $E(k) \sim k_x^{-3}$, $B^2(k) \sim k_x^{-9/2}$  & \\
      run-12 &  $Z(k) \sim k_x^{-2}$, $E(k) \sim k_x^{-3}$, $B^2(k) \sim k_x^{-9/2}$  & \\
      &    &\\
      run-13&  $Z(k) \sim k_x^{-2}$, $E(k) \sim k_x^{-3}$  &  \\
      run-14 & $Z(k) \sim k_x^{-2}$, $E(k) \sim k_x^{-7/2}$, $B^2(k) \sim k_x^{-3}$  & \\
      &  & &\\
      run-15 & $Z(k) \sim k_x^{-2}$, $E(k) \sim k_x^{-3}$, $B^2(k) \sim k_x^{-5}$  & \\
      run-16& $Z(k) \sim k_x^{-5/2}$, $E(k) \sim k_x^{-4}$, $B^2(k) \sim k_x^{-3}$  & \\
      &  & &\\
      run-17 & $Z(k) \sim k_x^{-2}$, $E(k) \sim k_x^{-9/2}$, $B^2(k) \sim k_x^{-3}$  & \\
      
    \end{tabular}
    \caption{Summary of power laws for the numerical simulations in
      inertial sub-range.}
    \label{Tab4}
  \end{center}
\end{table}


\subsection{Simulation results for constant viscosity, irresistive cases ($Pr_m = \infty$): run-3-6} \label{sec:constvisc}

Constant viscosity is introduced throughout the domain in the
simulation.  The simulations are performed for two different values of
constant Reynolds numbers, $Re= 2\times 10^3$ and $Re=2 \times 10^6$,
but with no resistivity ($Re_m=\infty$). As $\eta =0$ for these
simulations, this study corresponds to the cases of very large
magnetic Prandtl number ($Pr_m \rightarrow \infty$). In this study,
the case without magnetic field $\beta^{ini} \rightarrow \infty$ and
with magnetic field $\beta^{ini}=5000$ at $t\gamma_{RT} =0$ are
considered.  The relevant simulation parameters are shown in
Table~\ref{Tab2} under run-3-6.
\begin{figure*}
  \centering
  \includegraphics[width=\linewidth]{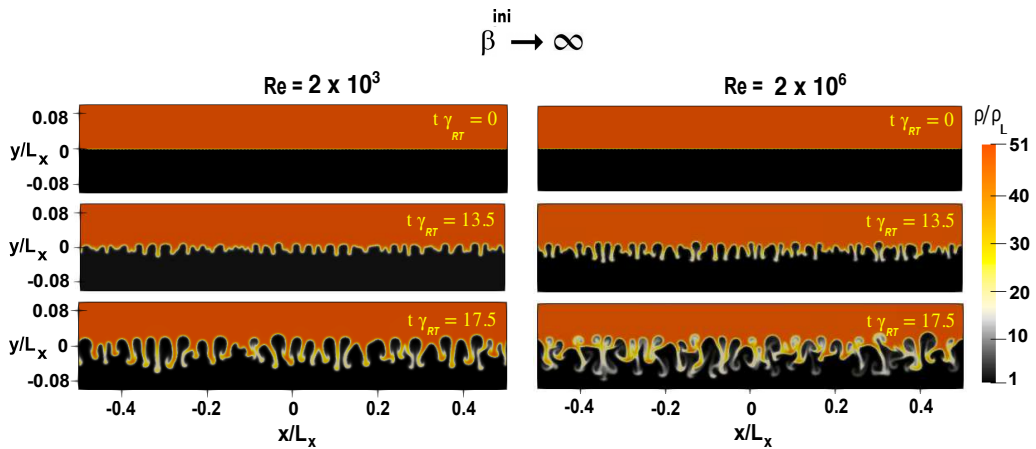}
  \caption{Plot of mass density ($\rho/\rho_L$) profile at different
    times for different constant values of $Re$; where $\beta^{ini}
    \rightarrow \infty$ and $Pr_m=\infty$ ($\eta=0$). Note the stabilizing effect of
    viscosity on short-wavelength RTI.}
  \label{fig6}
\end{figure*}
\begin{figure*}
  \centering
  \includegraphics[width=\linewidth]{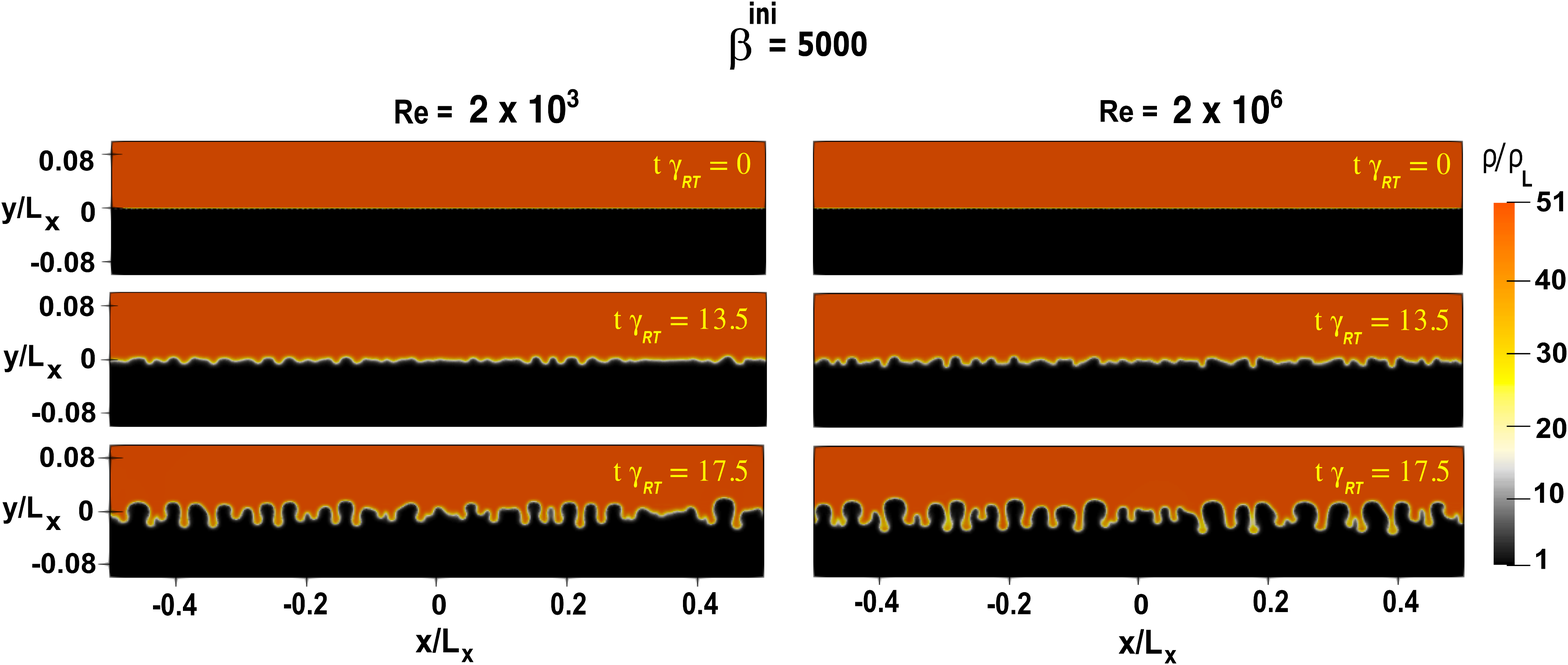}
  \caption{Plot of mass density ($\rho/\rho_L$) profile at different
    times for different constant values of $Re$; where $\beta^{ini}
    =5000$ and $Pr_m=\infty$ ($\eta=0$). Note the viscous and magnetic
    field stabilization acting in tandem to damp RTI growth.}
  \label{fig7}
\end{figure*}
In Fig.~\ref{fig6}, the mass density ($\rho/\rho_L$) is shown at
different times for $Re =2 \times 10^3$ and $Re =2\times 10^6$ for
$\beta^{ini} \rightarrow \infty$. It is seen that the growth of the
RTI decreases with decreasing $Re$ or increasing viscosity ($\mu$).
Fig.~\ref{fig7} shows mass density ($\rho/\rho_L$) at different times
for the two Reynolds numbers $Re =2 \times 10^3$ and $Re =2 \times
10^6$, but for $\beta^{ini}=5000$.  Here, the size of the RTI fingers
decreases further when applying a horizontal magnetic field compared
with the inviscid case presented in Section~\ref{sec:inviscirres}. The
magnetic field has a stabilizing effect in addition to viscous
stabilization on the growth of RTI. To further illustrate the
complementary role of viscous and magnetic field stabilization, the
peak bubble-to-spike distance ($h/L_x$) over time ($t \gamma_{RT}$) is
presented for both $\beta^{ini} \rightarrow \infty$ and
$\beta^{ini}=5000$ and for different constant Reynolds numbers ($Re$)
in Fig.~\ref{fig8}. Note that, as $Re$ increases the growth rate of
the RTI approaches the growth rate for the inviscid cases ($\mu=0$)
with and without the initial magnetic field.  For $\beta^{ini}
\rightarrow \infty$, the growth rate from the simulations is $0.55
\gamma_{RT}$ and $0.64 \gamma_{RT}$ for $Re =2 \times 10^3$ and $Re =2
\times 10^6$, respectively.  The analytical expression for the growth
rate of RTI in a compressible viscous fluid is given by
\cite{menikoff1977unstable},
\begin{equation}
  \gamma_{RT}^{vis} = \sqrt{A_t gk} \left(\sqrt{1+ \omega} -\sqrt{\omega} \right)
\end{equation} 
where $\omega= \bar{\nu}^2 k^3/ A_t g$ and $\bar{\nu} =(\mu_l +\mu_h)/
(\rho_l+ \rho_h)$ is the density averaged kinematic viscosity.  In
Fig~\ref{fig9}, the analytical form of $ \gamma_{RT}^{vis}/
\gamma_{RT} $ is shown as a function of wave number $k_x L_x$ for $Re
= 2 \times 10^3$ and $Re = 2 \times 10^6$.  For $Re = 2 \times 10^3$,
it is seen that the analytical growth rate is maximum for $k_x L_x
\approx 60 \pi$ which corresponds to a wavelength of approximately
$L_x /30$.  Similarly, for $Re = 2 \times 10^6$, the analytical growth
rate becomes maximum for $k_x L_x \approx 76 \pi$ or a wavelength of
approximately $L_x /38$.  This is consistent with the simulation
results from Fig.~\ref{fig7}.  The theoretical growth rate of the mode
having wavelength $L_x/30$ and for the mode having wavelength $L_x/38$
are approximately $0.56 \gamma_{RT}$ and $0.65 \gamma_{RT}$,
respectively.  The growth rates obtained from simulations show good
agreement with the analytical results.

\begin{figure}
  \centering
  \includegraphics[width=0.6\linewidth]{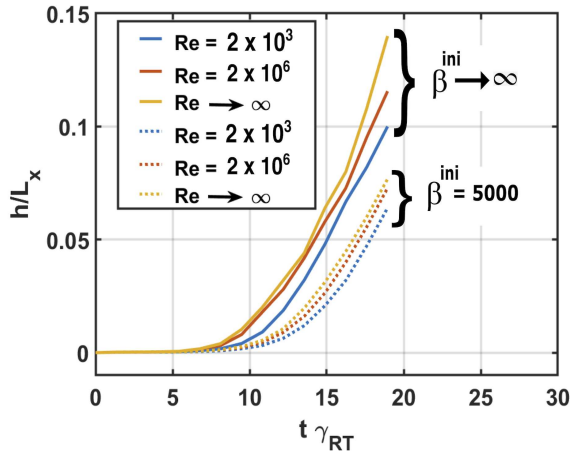}
  \caption{Plot of peak bubble-to-spike distance ($h/L_x$) over time
    ($t \gamma_{RT}$) for $\beta^{ini} \rightarrow \infty$ and
    $\beta^{ini} =5000$ and for different constant values of $Re$;
    where $\eta=0$.  Note the effects of viscous and magnetic
    stabilization on RTI growth.}
  \label{fig8}
\end{figure}

\begin{figure*}
  \centering
  \includegraphics[width=0.6\linewidth]{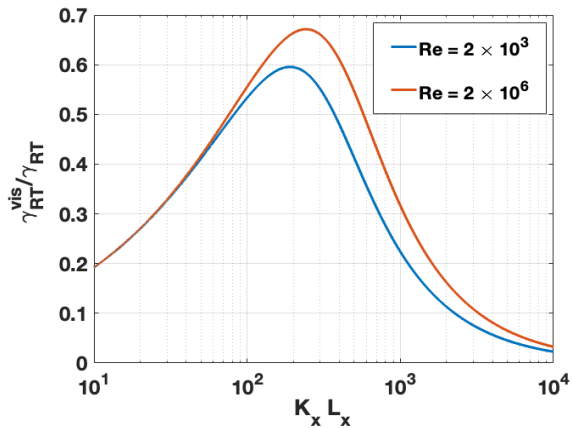}
  \caption{Plot of $\gamma_{RT}^{vis}/\gamma_{RT}$ as a function of
    wave number $k_x L_x$ for different constant values of $Re$; where
    $\eta=0$.  Note that the viscous cases produce a peak growth in
    the linear regime corresponding to $k_xL_x \approx 60\pi$.}
  \label{fig9}
\end{figure*}

Note that, when viscosity increases, the morphology of the RTI spikes
appear to be smooth and exhibit different characteristics as seen in
Fig.~\ref{fig7}. Due to the presence of viscosity, the traditional
mushroom cap structure on the tip of the RTI fingers gets inhibited
and forms smooth structures. The presence of viscosity also strongly
suppresses the growth of the small scale structures and
short-wavelength modes.

\begin{figure}
  \centering
  \includegraphics[width=0.6\linewidth]{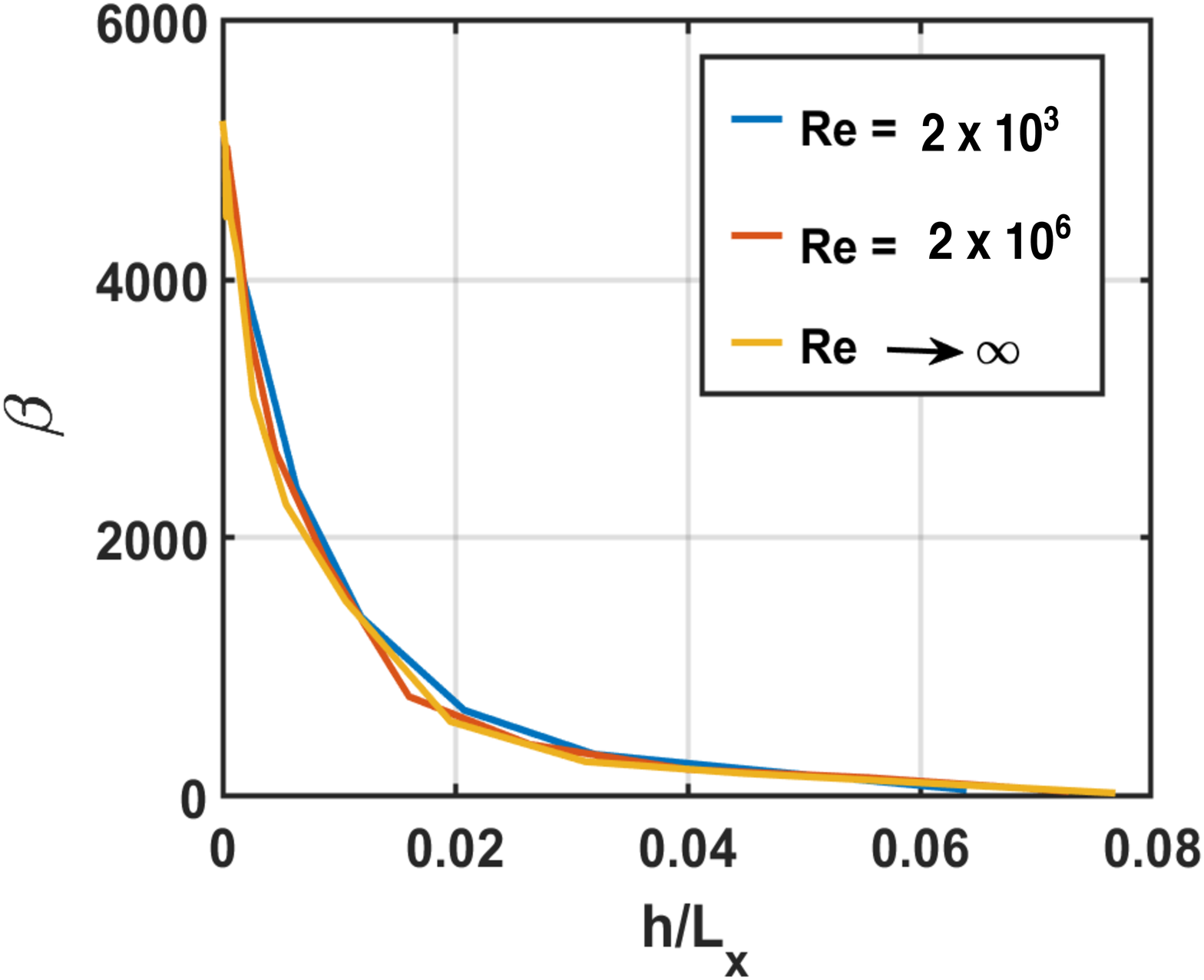}
  \caption{Plot of plasma $\beta$ as a function of peak
    bubble-to-spike distance ($h/L_x$) for different values of $Re$;
    where $\beta^{ini}=5000$ and $Pr_m=\infty$ ($\eta=0$).  Note that
    the plasma $\beta$ is independent of $Re$ when compared at the
    same RTI amplitude. }
  \label{fig10}
\end{figure}

\begin{figure*}
  \centering
  \includegraphics[width=\linewidth]{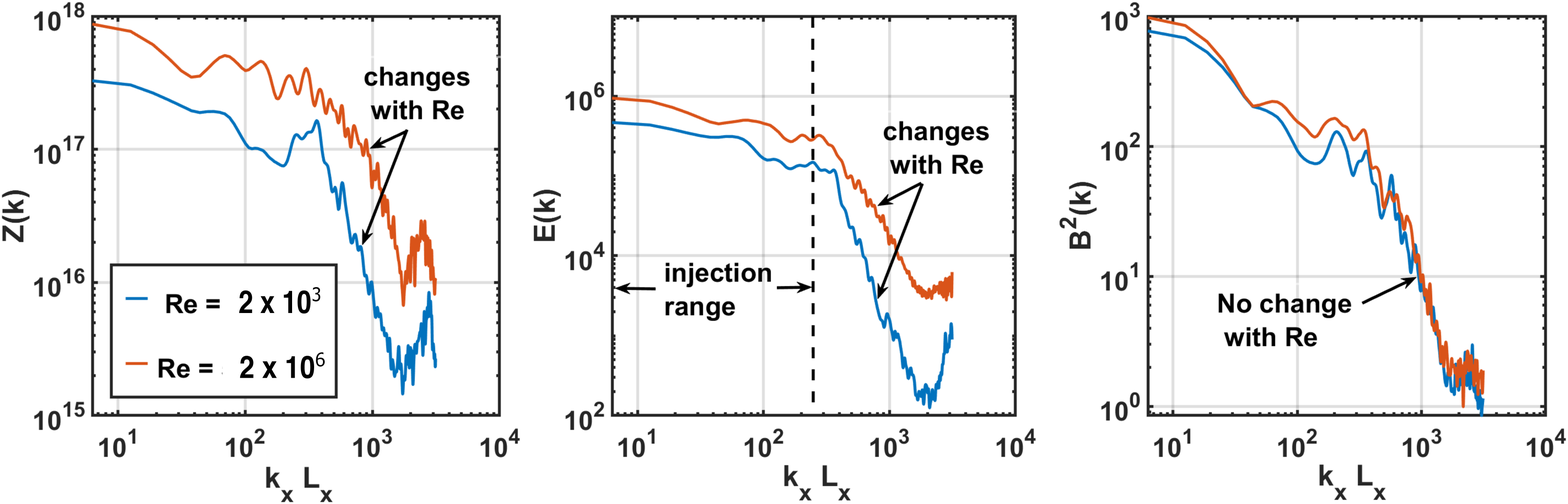}
  \caption{Evolution of enstrophy ($Z$), kinetic energy ($E$), and
    magnetic field energy ($B^2$) spectra at late-time $t\gamma_{RT} =
    17.5$ as a function of wave number ($k_x L_x$) for different
    constant values of $Re$; where $\eta=0$. Note that the spectra of
    enstrophy and kinetic energy in the inertial range of dissipation
    ($k_x L_x \geq 80\pi$) changes with $Re$ but the spectra of
    magnetic field in this range is independent of $Re$.}
  \label{fig11}
\end{figure*}

The plasma $\beta$ as a function of peak bubble-to-spike distance
($h/L_x$) for different $Re$ for $\beta^{ini}=5000$ is presented in
Fig.~\ref{fig10}. Note that plasma $\beta$ is independent of $Re$ if
presented as a function of the peak bubble-to-spike amplitude instead
of as a function of time. This shows that the dynamics of magnetic
field is not affected by the viscosity for the same amplitude of the
RTI growth but the actual RTI growth as a function of time is impacted
by the different $Re$ as noted from Fig.~\ref{fig8}. Also note that
plasma $\beta$ decreases with time or height as RTI grows for all $Re$
considered. This is because the value of magnetic field increases as
RTI grows in the system.  Figure~\ref{fig11} presents enstrophy($Z$),
kinetic energy ($E$) and magnetic field energy ($B^2(k)$) spectra at
time $t \gamma_{RT} t= 17.5$ as a function of wave number $k_x L_x$
for different values of $Re$. The scaling of these power laws in both
injection and inertial sub-range for these cases (run-3-6) have been
summarized in Tables~\ref{Tab3} and \ref{Tab4}.  Note that the
spectral power of the magnetic energy does not change with $Re$ but
the spectral power of enstrophy and kinetic energy increases with
increasing the value of $Re$ for all available modes. This shows that
the dynamics of magnetic field energy is independent of $Re$ or
viscosity. It is shown by \cite{Kulsrud} that the dynamics of the
magnetic field can be completely described by ion fluid vorticity in
the absence of viscosity and resistivity but in the presence of a
Biermann battery, which is not considered in this work. Including the
viscosity and resistivity into the MHD equations considered here, a
theoretical treatment is included to illustrate the dynamics of
magnetic field and vorticity in presence of viscosity and resistivity.
Following the same method as shown by Kulsrud et al. \cite{Kulsrud},
the momentum equation (Eq. \ref{eq2}) can be written in terms of
vorticity ($\omega$) as,
\begin{equation}
  \frac{\partial \mathbf{\omega}}{\partial t} =\frac{\mathbf{\nabla}\rho \times \mathbf{\nabla}P}{\rho^2} +\mathbf{\nabla} \times (\mathbf{u} \times \mathbf{\omega} ) +\mathbf{\nabla} \times \frac{\mathbf{J} \times \mathbf{B}}{\rho}   -\mathbf{\nabla} \times \frac{\mathbf{\nabla} \cdot \mathbf{\pi}}{\rho}
  \label{eq6}
\end{equation}
where $\mathbf{J}$ represents the net current density. Similarly,
Eq. (\ref{eq4}) can be modified in terms of ion cyclotron frequency
($\mathbf{\omega_{ci}}=Z_i e \mathbf{B}/m_i$) as,
\begin{equation}
  \frac{\partial \mathbf{\omega_{ci}}}{\partial t} = \nabla \times (\mathbf{u} \times \mathbf{\omega_{ci}} ) - {\frac{1}{\mu_0} \nabla \times (\eta \nabla \times \mathbf{\omega_{ci}})}.
  \label{eq7}
\end{equation}
The last term on right hand side of equations (\ref{eq6}) and
(\ref{eq7}) are responsible for the dissipation of the vorticity and
magnetic field, respectively. The dynamics of vorticity and kinetic
energy depend on the viscous stress tensor $\mathbf{\pi}$ and the
corresponding $Re$.  This is consistent with the numerical results
presented here.  On the other hand, the dynamics of vorticity is
independent of resistivity $\eta$ or magnetic Reynolds number $Re_m$,
but the dynamics of magnetic field depends on the $Re_m$. To
illustrate this, simulations are performed for different constant
values of $Re_m$ discussed in the next section.


\subsection{Simulation results for constant resistivity, inviscid cases ($Pr_m=0$): run-7-8} \label{sec:constres}

\begin{figure*}
  \centering \includegraphics[width=\linewidth]{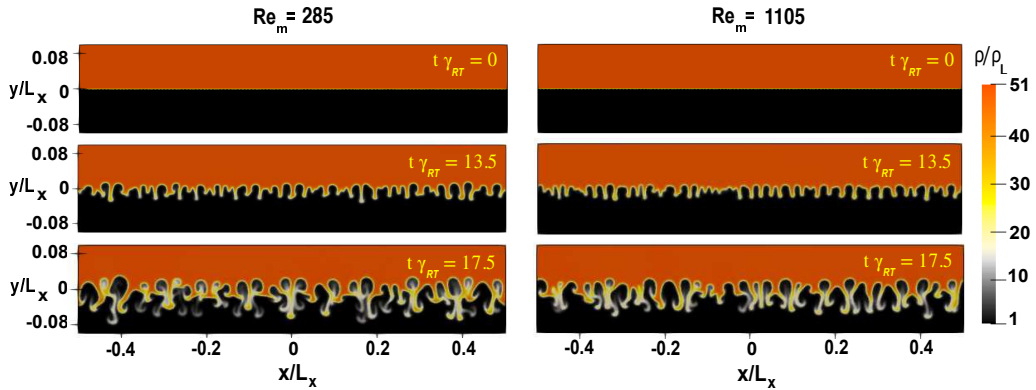}
  \caption{Plot of mass density ($\rho/\rho_L$) profile at different
    times for $Re_m=285$ and $Re_m=1105$; where $\beta^{ini}=5000$ and
    $\mu=0$. Note that a smaller $Re_m$ corresponding to a larger
    $\eta$ produces an increase in RTI growth compared to a larger
    $Re_m$.}
  \label{fig12}
\end{figure*}
\begin{figure}
  \centering
  \includegraphics[width=0.6\linewidth]{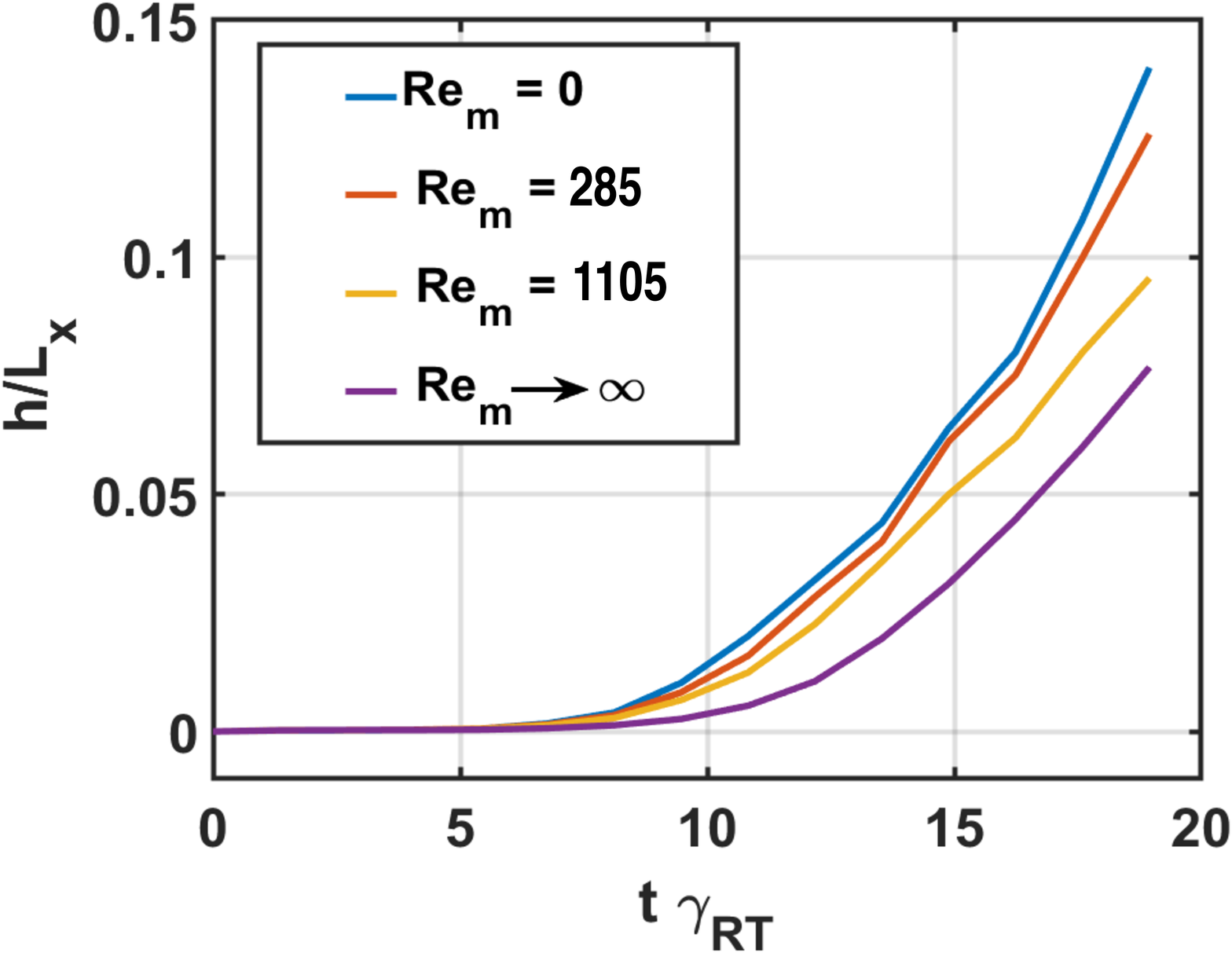}
  \caption{Plot of peak bubble-to-spike distance ($h/L_x$) over time
    ($t \gamma_{RT}$) for different constant values of $Re_m$; where
    $\beta^{ini}=5000$ and $\mu=0$.  Note that the growth rate
    increases with a decrease in $Re_m$ corresponding to an increase
    in $\eta$.}
  \label{fig13}
\end{figure}

In this section, simulation results are presented for different
constant magnetic Reynolds numbers ($Re_m$) but with no viscosity
($\mu=0$) (see run 7-8 in Table \ref{Tab2}).  In this study, $Pr_m
=0$. In all these simulations, an initial horizontal magnetic field
with $\beta^{ini} =5000$ is applied. In Fig.~\ref{fig12}, the mass
density ($\rho/\rho_L$) profile at different times is presented for
$Re_m =285$ and $Re_m=1105$. It is seen that the growth of the RTI
increases with a decrease in magnetic Reynolds numbers ($Re_m$) or
increase of resistivity ($\eta$). This is because the resistivity
diffuses the magnetic fields and reduces the magnetic
stabilization. As a result, the RTI growth increases due the reduction
of effective magnetic field tension. In this figure, it is to be noted
that the morphology of the RTI spikes in terms of mushroom cap
structures on the tip of the fingers are seen to be independent of
$Re_m$. Also of note is the appearance of additional small scale
structures for higher resistivity cases. This is also expected as the
magnetic field opposes development of the small scale structures. In
Fig.~\ref{fig13}, the peak bubble-to-spike distance ($h/L_x$) over
time ($t \gamma_{RT}$) is presented for different constant values of
$Re_m$ to illustrate the effect of magnetic Reynolds number on the
growth rate of RTI in HED plasmas. It is found that the growth rate
increases with increase in resistivity.  The numerical growth rates
are obtained from the simulations for $Re_m =285$ and $Re_m =1105$ as
$0.68 \gamma_{RT}$ and $0.53 \gamma_{RT}$, respectively.  Including a
finite constant resistivity $\eta$, \cite{jukes1963rayleigh} has shown
that the analytical growth rate of RTI changes with resistivity $\eta$
as,
\begin{equation}
  \gamma_{RT}^{res} \propto \eta^{1/3}.
\end{equation} 
The growth rates obtained from the simulations also obey the
analytical scaling.

\begin{figure}
  \centering
  \includegraphics[width=0.6\linewidth]{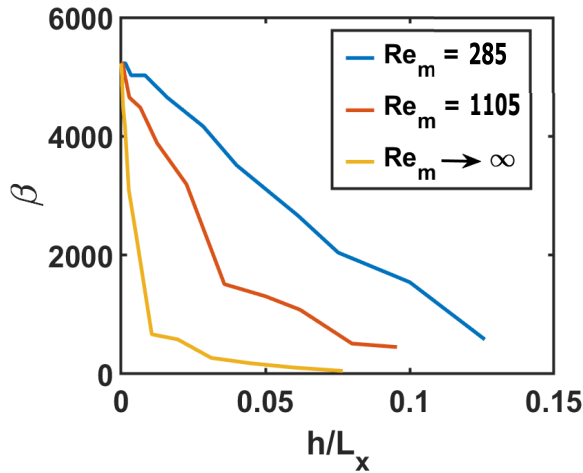}
  \caption{Plot of plasma $\beta$ as a function of peak
    bubble-to-spike distance ($h/L_x$) for different constant values
    of $Re_m$; where $\beta^{ini}=5000$ and with $\mu=0$. Note that
    the plasma $\beta$ changes with $Re_m$.}
  \label{fig14}
\end{figure}

The plasma $\beta$ is plotted as a function of peak bubble-to-spike
distance ($h/L_x$) for different $Re_m$ in Fig.~\ref{fig14}. Note that
plasma $\beta$ decreases with peak bubble-to-spike distance for all
values of $Re_m$ but at different rates depending on the value of
$Re_m$. The rate at which the plasma beta decreases is larger for high
$Re_m$. This shows that the dynamics of the magnetic field is not
independent of resistivity.  This is due to the fact that the magnetic
field gets diffused more for low $Re_m$ leading to a higher plasma
$\beta$.

\begin{figure*}
  \centering
  \includegraphics[width=\linewidth]{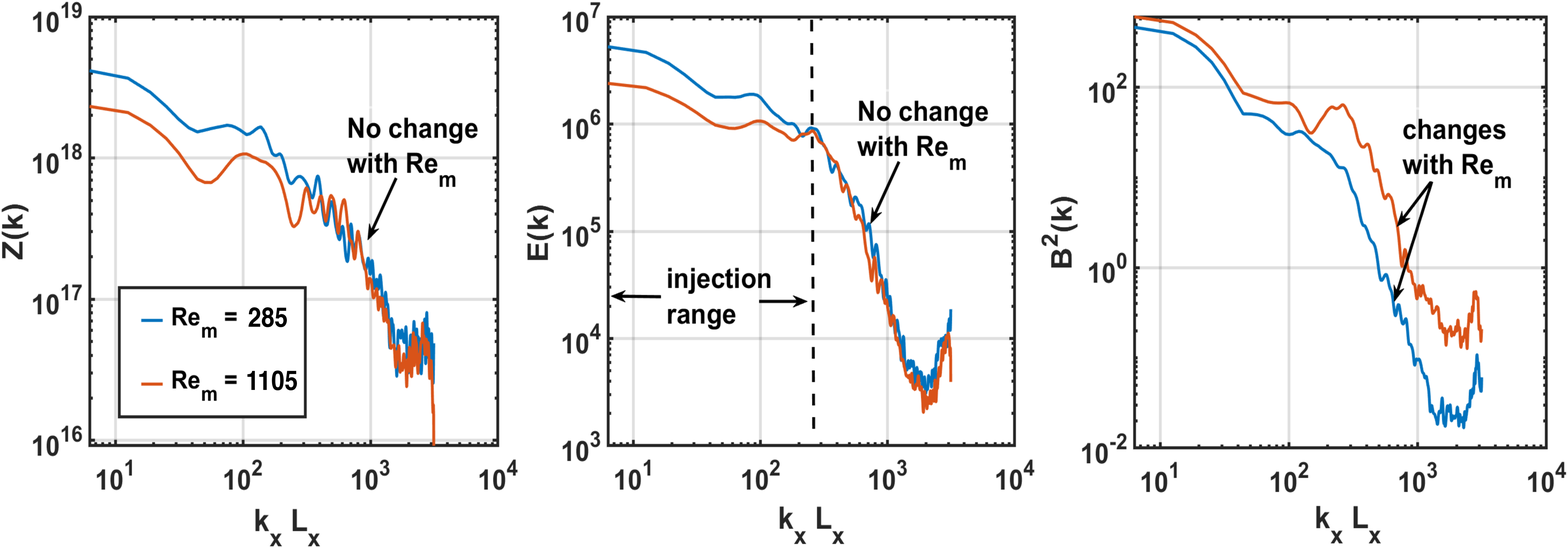}
  \caption{Evolution of enstrophy ($Z$), kinetic energy ($E$), and
    magnetic field energy ($B^2$) spectra at late-time $t\gamma_{RT} =
    17.5$ as a function of wave number ($k_x L_x$) for different
    constant values of $Re_m$; where $\beta^{ini}=5000$ and
    $\mu=0$. Note that the spectra of enstrophy and kinetic energy in
    the inertial range of dissipation ($k_x L_x \geq 80\pi$) are
    independent of $Re_m$ but the spectra of magnetic field in this
    range changes with $Re_m$.}
  \label{fig15}
\end{figure*}
In Fig.~\ref{fig15}, the plot of enstrophy($Z$), kinetic energy ($E$),
and magnetic field energy ($B^2(k)$) spectra at time ($t \gamma_{RT}
t= 17.5$) as a function of wave number $k_x L_x$ has been shown for
different values of $Re_m$.  The scaling of these power laws in both
injection and inertial sub-range for these cases (run-7-8) have been
summarized in Tables~\ref{Tab3} and \ref{Tab4}.  It is observed that
the magnetic field spectra changes significantly by changing the value
of $Re_m$, whereas the spectra of enstrophy and kinetic energy does
not show any significant dependence on the value of $Re_m$. The
spectral power of magnetic field energy increases with increasing the
value of $Re_m$ for all the available modes. This justifies that the
dynamics of magnetic field energy depends on $Re_m$ or $\eta$. But the
dynamics of enstrophy and kinetic energy does not depend on $Re_m$.
This is consistent with equations (\ref{eq6}) and (\ref{eq7}).

\subsection{Simulation results for constant viscosity, constant resistivity cases: run-9-12}

\begin{figure*}
  \centering
  \includegraphics[width=\linewidth]{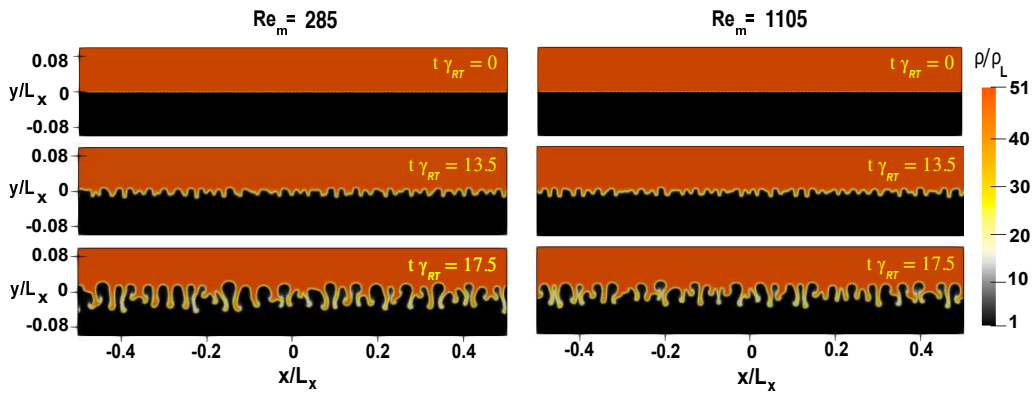}
  \caption{Plot of mass density ($\rho/\rho_L$) profile at different
    times for $Re_m=285$ ($Pr_m = 0.1$) and $Re_m=1105$ ($Pr_m =
    0.5$); where $Re= 2 \times 10^3$ and $\beta^{ini}=5000$.}
  \label{fig16}
\end{figure*}

\begin{figure*}
  \centering
  \includegraphics[width=\linewidth]{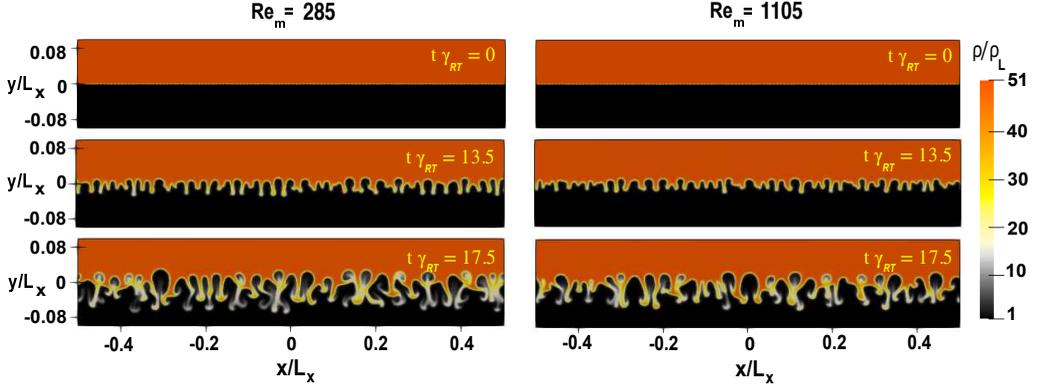}
  \caption{Plot of mass density ($\rho/\rho_L$) profile at different
    times for $Re_m=285$ ($Pr_m = 1 \times 10^{-4}$) and $Re_m=1105$
    ($Pr_m = 5 \times 10^{-4}$); where $Re= 2 \times 10^6$ and
    $\beta^{ini}=5000$.}
  \label{fig17}
\end{figure*}

Simulations have also been performed for different values of constant
$Re$ with the inclusion of different constant values of $Re_m$ (see
run-9-12 in Table~\ref{Tab2}). In this case, all the simulations use
an applied horizontal magnetic field corresponding to
$\beta^{ini}=5000$.  Fig.~\ref{fig16} presents the mass density
($\rho/\rho_L$) profile at different times for $Re_m =285$ ($Pr_m =
0.1$) and $Re_m=1105$ ($Pr_m = 0.5$) with $Re= 2 \times 10^3$.
Similarly, the mass density ($\rho/\rho_L$) profile at different times
for $Re_m =285$ ($Pr_m = 1 \times 10^{-4}$) and $Re_m=1105$ ($Pr_m = 5
\times 10^{-4}$) for $Re= 2 \times 10^6$ is presented in
Fig.~\ref{fig17}. Note that the morphology of the RTI fingers doesn't
exhibit a strong dependence on $Re_m$ for the values considered here,
but shows more significant dependence with $Re$. The mushroom caps on
the tip of the RTI fingers are inhibited for high viscosity. When
viscosity is held constant, the growth rate increases with an increase
in resistivity.  On the other hand, the growth rate decreases with an
increase in viscosity when resistivity is held constant.

\begin{figure*}
  \centering
  \includegraphics[width=\linewidth]{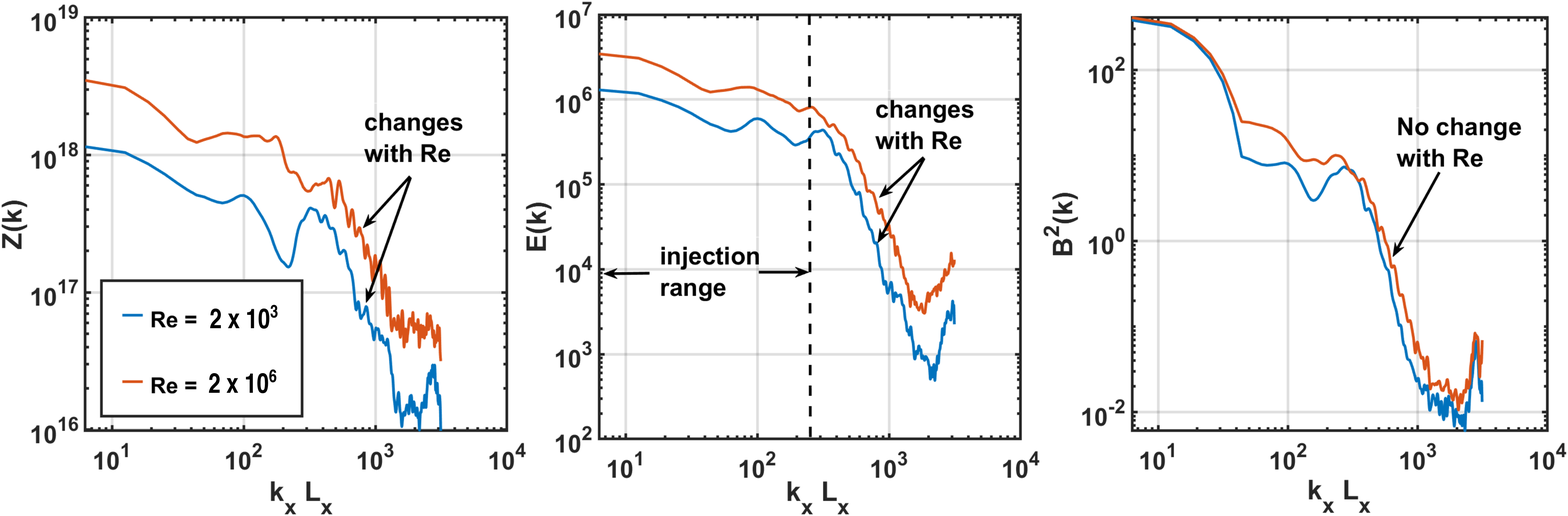}
  \caption{Evolution of enstrophy ($Z$), kinetic energy ($E$), and magnetic field energy ($B^2$) spectra at late-time $t\gamma_{RT} = 17.5$ as a function of wave number ($k_x L_x$)  for different constant values of $Re$; where $\beta^{ini}=5000$ and $Re_m =285$. Note that the spectra of enstrophy and kinetic energy in the inertial range of dissipation ($k_x L_x \geq 80\pi$) change with $Re$ but the spectra of magnetic field in this range is independent of $Re$.}
  \label{fig18}
\end{figure*}

\begin{figure*}
  \centering
  \includegraphics[width=\linewidth]{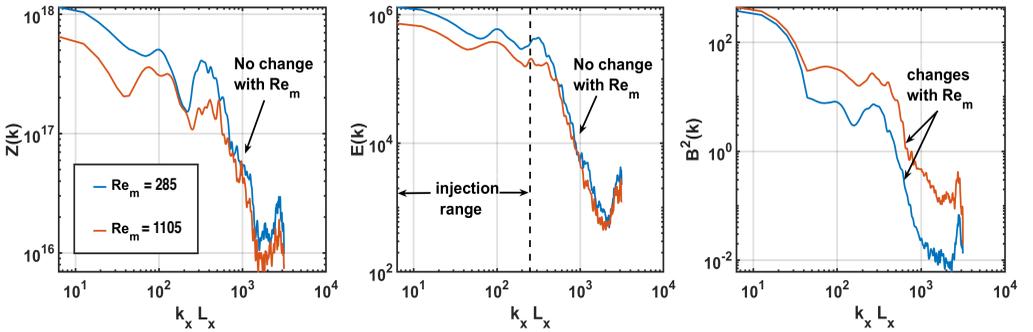}
  \caption{Evolution of enstrophy ($Z$), kinetic energy ($E$), and magnetic field energy ($B^2$) spectra at late-time $t\gamma_{RT} = 17.5$ as a function of wave number ($k_x L_x$)  for different constant values of $Re_m$; where  $\beta^{ini}=5000$ and $Re = 2  \times 10^3$. Note that the spectra of enstrophy and kinetic energy in the inertial range of dissipation ($k_x L_x \geq 80\pi$) are independent of $Re_m$ but the spectra of magnetic field changes with $Re_m$.}
  \label{fig19}
\end{figure*}

The power law scaling of the enstrophy ($Z$), kinetic energy ($E$),
and magnetic field energy ($B^2(k)$) spectra as a function of wave
numbers $k_x L_x$ is quantified for these runs (run-9-12) in both the
injection range as well as the inertial sub-range.  The scalings of
these power laws are given in Tables~\ref{Tab3} and \ref{Tab4} for
run-9-12.  Fig.~\ref{fig18} presents the enstrophy($Z$), kinetic
energy ($E$) and magnetic field energy ($B^2(k)$) spectra at time $t
\gamma_{RT} t= 17.5$ as a function $k_x L_x$ for different values of
$Re$; where the value of $Re_m$ is held constant to $Re_m=285$ for
$\beta^{ini}=5000$.  It is seen here that the spectra of enstrophy and
kinetic energy change with $Re$, whereas the magnetic field spectra
does not change with $Re$. Similarly, the enstrophy($Z$), kinetic
energy ($E$), and magnetic field energy ($B^2(k)$) spectra at time $t
\gamma_{RT} t= 17.5$ as a function $k_x L_x$ for different values of
$Re_m$ are plotted in Fig.~\ref{fig19} holding $Re$ constant at $Re=2
\times 10^3$ for $\beta^{ini}=5000$. Note that $Re_m$ does not affect
the spectra of enstrophy and kinetic energy, whereas the magnetic
field spectra depends on $Re_m$.  These findings are consistent with
those in Sections \ref{sec:constvisc} and \ref{sec:constres}.


\subsection{Simulation results for fully varying viscosity, irresistive cases ($Pr_m=\infty$): run-13-14}

\begin{figure*}
  \centering
  \includegraphics[width=\linewidth]{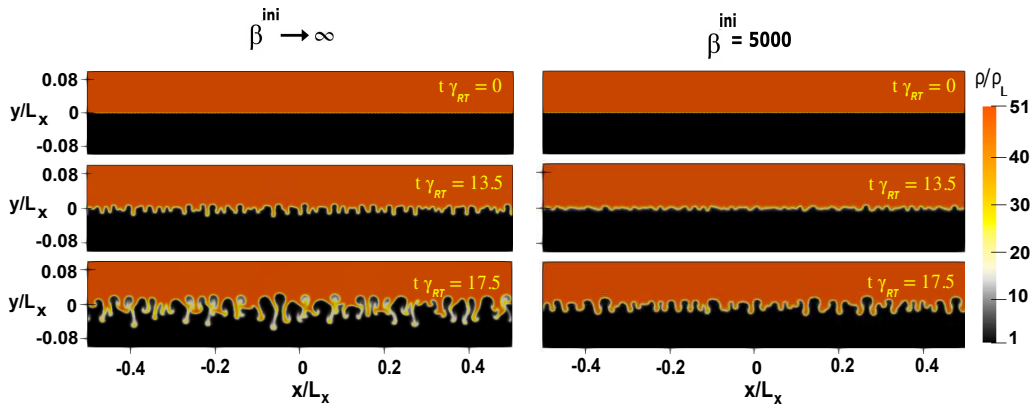}
  \caption{Plot of mass density ($\rho/\rho_L$) profile at different
    times for $\beta^{ini} \rightarrow \infty$ and $\beta^{ini}=5000$;
    where fully varying $Re$ is considered with no resistivity
    ($\eta=0$).  Note the viscous stabilization of RTI due to the high
    viscosity of the low density region. }
  \label{fig20}
\end{figure*}

\begin{figure}
  \centering
  \includegraphics[width=0.6\linewidth]{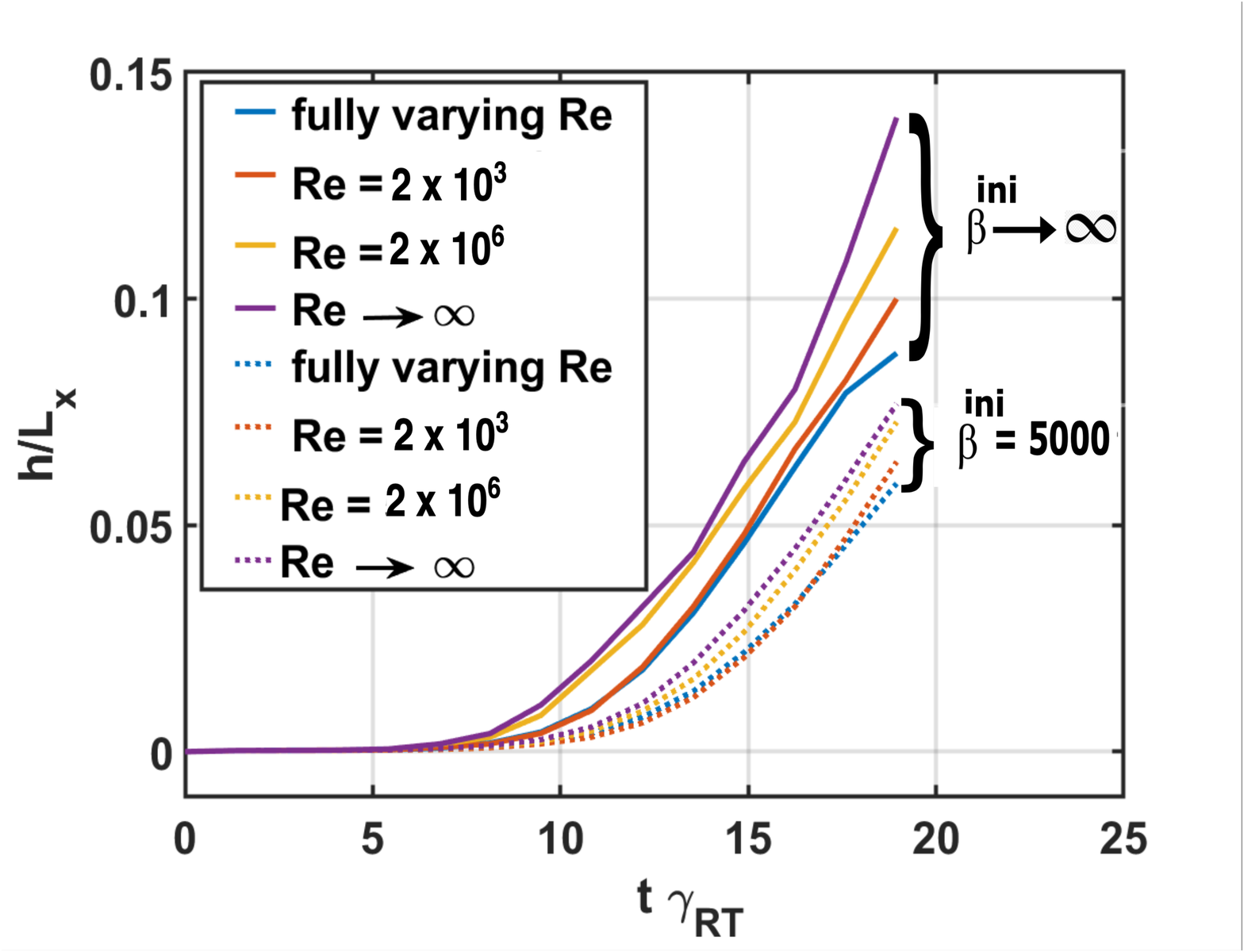}
  \caption{Plot of peak bubble-to-spike distance ($h/L_x$) over time
    ($t \gamma_{RT}$) for $\beta^{ini} \rightarrow \infty$ and
    $\beta^{ini}=5000$ for different values of $Re$; where $\eta=0$.
    Note that the fully varying $Re$ case has viscous stabilization
    corresponding to the viscosity of the lower density fluid.}
  \label{fig21}
\end{figure}

Next, the self-consistent fully varying $Re$ profile shown in
Fig.~\ref{fig1} is considered without resistivity (see run-13-14 in
Table~\ref{Tab2}). The simulations have been performed using both
$\beta^{ini} \rightarrow \infty$ and $\beta^{ini}=5000$.  In
Fig.~\ref{fig20}, the mass density ($\rho/\rho_L$) profile is
presented at different times for $\beta^{ini} \rightarrow \infty$ and
$\beta^{ini}=5000$.  To further illustrate the effect of a fully
varying $Re$ profile on the RTI, the peak bubble-to-spike distance
($h/L_x$) over time ($t \gamma_{RT}$) is presented for $\beta^{ini}
\rightarrow \infty$ and $\beta^{ini}=5000$ for this case in
Fig.~\ref{fig21} along with the bubble-to-spike amplitudes for
constant $Re$ cases. The growth and nature of the RTI for fully
varying viscosity for $\beta^{ini} \rightarrow \infty$ and
$\beta^{ini}=5000$ is close to that of the high viscosity case or low
Reynolds number ($Re=2 \times 10^3$) case. This is because, the RTI
fingers largely grow in the lower density regime ($y<0$) at the
interface due to the high Atwood number considered here.  The mixing
is not significant in the high density regime. In the lower density
regime, the value of $Re=2 \times 10^3$, which has significantly
higher viscosity compared to the high density regime
($y>0$). Therefore, the evolution of RTI is dominated by the high
viscosity regime.  Hence, viscosity, even if disparate, plays an
important role in the RTI process in such parameter regimes with and
without an applied horizontal magnetic field.  Similar to the previous
cases, the power law scaling of enstrophy ($Z$), kinetic energy ($E$),
and magnetic field energy ($B^2$) spectra as a function of wave number
$k_x L_x$ in both injection and inertial sub-range are summarized in
Tables~\ref{Tab3} and \ref{Tab4} under run-13-14.



\subsection{Simulation results for fully varying viscosity, constant resistivity cases: run-15-16}
\begin{figure*}
  \centering
  \includegraphics[width=\linewidth]{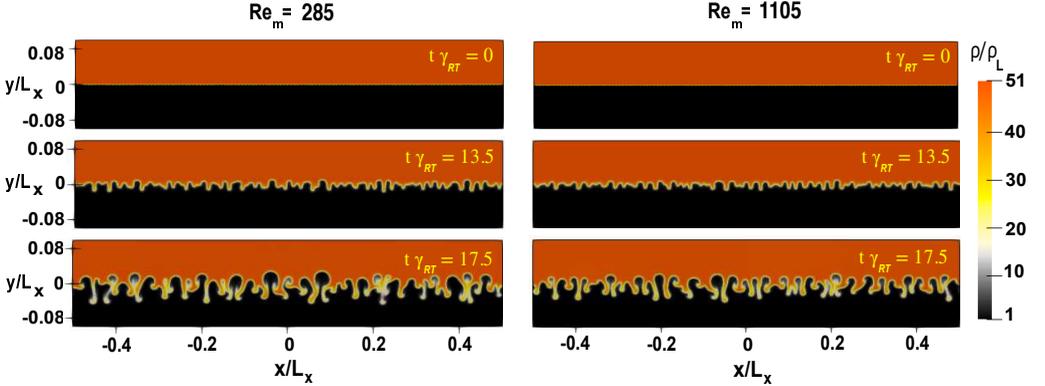}
  \caption{Plot of mass density ($\rho/\rho_L$) profiles at different
    times for $Re_m =285$ and $Re_m= 1105$; where $\beta^{ini}=5000$
    and fully varying $Re$ is considered.  Note increased growth of
    RTI for lower $Re_m$ as expected even with a fully varying $Re$.}
  \label{fig22}
\end{figure*}

\begin{figure}
  \centering
  \includegraphics[width=0.6\linewidth]{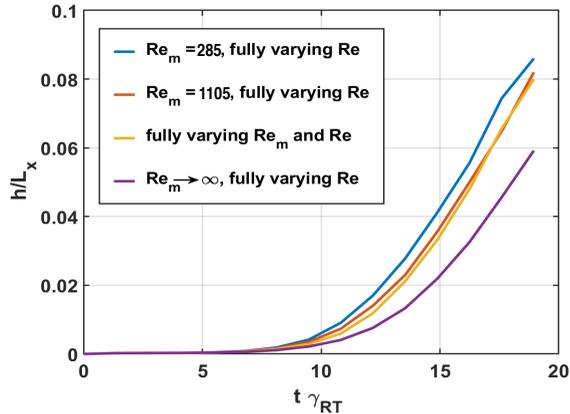}
  \caption{Plot of peak bubble-to-spike distance ($h/L_x$) over time
    ($t \gamma_{RT}$) for different values of $Re_m$; where
    $\beta^{ini}=5000$ and fully varying $Re$ is considered.  Note
    that RTI growth is higher for low $Re_m$ even with a fully varying
    $Re$.  Also note that the fully varying $Re$ and fully varying
    $Re_m$ case asymptotes to the $Re_m$ corresponding to the lower
    fluid.}
  \label{fig23}
\end{figure}

\begin{figure*}
	\centering
	\includegraphics[width=0.7\linewidth]{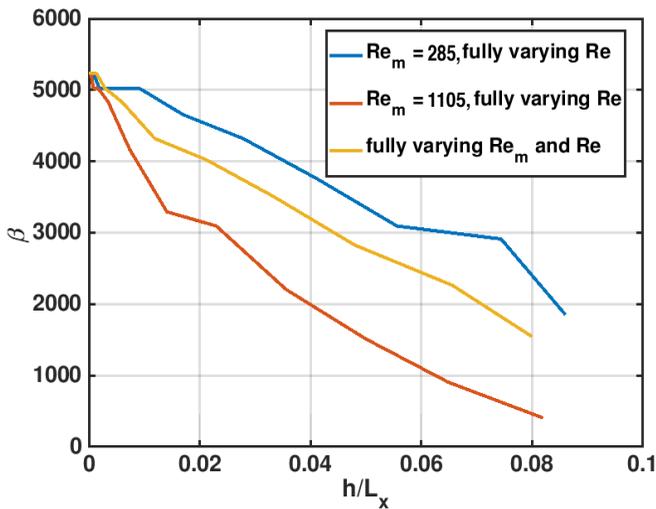}
	\caption{Plot of plasma $\beta$ as a function of peak
          bubble-to-spike distance ($h/L_x$) for $Re_m=285$, $Re_m
          =1105$, and fully varying $Re_m$; where $\beta^{ini}=5000$
          and fully varying $Re$ are considered.  Note that the plasma
          $\beta$ is higher late in time for lower $Re_m$ compared
          with a higher $Re_m$ even with a fully varying $Re$. Also
          note that the fully varying $Re_m$ case produces a magnetic
          field that lies in between the regimes of the upper and
          lower fluids.}
	\label{fig24}
\end{figure*}

Simulations are performed considering fully varying $Re$ with the
inclusion of different values of constant $Re_m$ ($Re_m=285$ and
$Re_m=1105$).  These correspond to $Pr_m$ ranging from $0.5 - 4\times
10^{-6}$ for $Re_m=285$ and $Pr_m$ ranging from $2 - 1.5\times
10^{-5}$ for $Re_m = 1105$.  In these studies, an applied horizontal
magnetic field corresponding to $\beta^{ini}=5000$ is included as
before. Fig.~\ref{fig22} shows the mass density ($\rho/\rho_L$)
profile at different times for $Re_m =285$ and $Re_m=1105$. It is seen
that the growth of the RTI spikes increases with the decrease of
$Re_m$ as expected.  In Fig.~\ref{fig23}, the peak bubble-to-spike
distance ($h/L_x$) over time ($t\gamma_{RT}$) for different values of
$Re_m$ is shown.  Note that the growth of RTI is higher for high
resistivity (blue solid line) compared to that obtained for low
resistivity (red solid line) when also including the fully varying
viscosity.  The plasma $\beta$ as a function of peak bubble-to-spike
distance ($h/L_x$) for different values of constant $Re_m$ (solid blue
and red line) is presented in Fig.~\ref{fig24}, where
$\beta^{ini}=5000$ and fully varying $Re$ are considered. The magnetic
field decreases for the lower value of $Re_m= 285$ which corresponds
to higher $\eta$.  Furthermore, it is observed here that the
morphology of the RTI fingers are not significantly affected by the
resistivity. The power law scaling of enstrophy ($Z$), kinetic energy
($E$), and magnetic field energy ($B^2$) spectra as a function of wave
number $k_x L_x$ in both injection and inertial sub-range for these
cases is summarized in Table~\ref{Tab3} and \ref{Tab4} in the column
under run-15-16.

\subsection{Simulation results for fully varying viscosity, fully-varying resistivity case: run-17}
\begin{figure*}
 	\centering
   \includegraphics[width=0.7\linewidth]{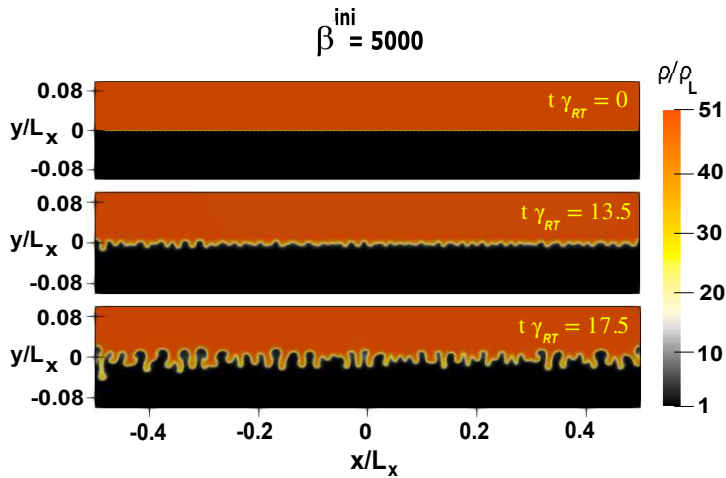}
   \caption{Plot of mass density ($\rho/\rho_L$) profiles at different
     times for fully varying viscosity and resistivity case; where
     $\beta^{ini}=5000$.  Note the morphology of the RTI in this case
     exhibiting less Kelvin-Helmholtz formation than even the higher
     $Re_m=1105$, fully varying $Re$ case presented in
     Fig.~\ref{fig22}.}
   \label{fig25}
 \end{figure*}

The final set of simulations are performed for a fully varying $Re$
along with a fully varying $Re_m$ profile.  These correspond to $Pr_m$
ranging from $2 - 4\times 10^{-6}$.  Note that the resistivity profile
used for this case is the modified resistivity profile shown in
Fig. \ref{fig1}. In this case, an applied horizontal magnetic field
corresponding to $\beta^{ini}=5000$ is included as in the previous
cases.  Fig.~\ref{fig25} presents the mass density ($\rho/\rho_L$)
profiles for fully varying viscosity and resistivity profiles at
different times. It is observed that the structure of RTI is quite
different from the conventional mushroom cap structure. The morphology
of the RTI spikes exhibits less Kelvin-Helmholtz formation and shows
the suppression of small scale structures more significantly than the
higher $Re_m=1105$, fully varying $Re$ case presented in
Fig.~\ref{fig22}.  In Fig.~\ref{fig23}, the peak bubble-to-spike
distance ($h/L_x$) over time($t\gamma_{RT}$) for fully varying $Re_m$
and fully varying $Re$ profiles is presented along with the constant
$Re_m$ cases (see yellow solid line).  The growth rate for the fully
varying resistivity case is close to the the growth rate obtained for
the constant $Re_m =1105$ case.  This is because the RTI mostly grows
in the low density regime where $Re_m =1105$. Therefore, the dynamics
of RTI for the high Atwood number regime can be described by the
physical parameter space of the lower fluid, which is governed by the
viscosity and resistivity of the lower fluid.  The plasma $\beta$ as a
function of peak bubble-to-spike distance ($h/L_x$) for fully varying
$Re_m$ and $Re$ is shown in Fig.~\ref{fig24} (see solid yellow line),
where $\beta^{ini}=5000$ is considered. The dynamics of the magnetic
field and its corresponding growth, as noted by the decreasing plasma
$\beta$, for fully varying $Re_m$ and $Re$ is different from the
constant magnetic $Re_m$ cases.  The field strength obtained lies
inbetween the regimes of the upper and lower fluid (with their
corresponding resistivities).  The power law scaling of enstrophy
($Z$), kinetic energy ($E$), and magnetic field energy ($B^2$) spectra
as a function of wave number $k_x L_x$ in both injection and inertial
sub-range is summarized in Table~\ref{Tab3} and \ref{Tab4} under run
17.

\section{Summary and conclusion}
\label{sec5}
In summary, the role of viscosity and resistivity on Rayleigh-Taylor
and magneto-Rayleigh-Taylor instabilities is studied for a high Atwood
number and high plasma-$\beta$ regime in high-energy-density (HED)
plasmas applicable to both laboratory and astrophysical settings.
This work describes 2D RTI evolution and resulting turbulence when
surveying plasma-$\beta$ and magnetic Prandtl number, $Pr_m$, for
these regimes.  The simulations have been performed using fluid
simulation techniques based on the unstructured discontinuous Galerkin
finite element method
\cite{song2020unstructured,song2021affine,hesthaven2007nodal}. Using a
visco-resistive-magnetohydrodynamic (MHD) model, a detailed
investigation of RTI in 2D planar geometry for experimentally and
observationally relevant parameters is presented. It has been shown
here that the inclusion of viscosity and resistivity in the system
drastically changes the growth of the instability as well as modifies
its internal structure on smaller scales.  The presence of viscosity
inhibits the development of small scale structures and significantly
modifies the morphology of the RTI spikes. On the other hand, the
morphology of the RTI spikes is found to be independent of resistivity
but it assists in the development of small scale structures via the
diffusion of the magnetic fields.  The reduced magnetic field strength
that results in time permits shorter wavelength modes to grow.
Considering fully varying viscosity and fully varying resistivity
profiles in the simulation due to the strong dependence of viscosity
and resistivity on the disparate temperature profile across the
interface, the effect of both viscosity and resistivity is shown to be
significant on the evolution of RTI in HED plasmas.  Furthermore, it
is also found that the dynamics of the magnetic field is explicitly
independent of viscosity and likewise the resistivity does not affect
the dynamics of enstrophy and kinetic energy.  Also presented here is
the power law scaling of enstrophy, kinetic energy, and magnetic field
energy over a wide range of viscosity and resistivity in both
injection range and inertial sub-range of spectra. This could provide
a useful tool for understanding RTI induced turbulent mixing in high
Atwood number HED plasmas and could aid in interpretation of
observations of RTI-induced turbulence spectra.

\section*{Acknowledgments}
 All simulations in this paper were performed using the Advanced
 Research Computing (ARC) at Virginia Tech.
  
\section*{Funding}
This work was supported by the National Science Foundation under grant
number PHY-1847905.

\section*{Declaration of interests}
The authors report no conflict of interest.

\bibliographystyle{jpp}

\bibliography{References}

\end{document}